\pdfoutput=1

\RequirePackage{fix-cm}
\RequirePackage{rotating}

\documentclass[smallextended]{svjour3}
\smartqed 

\usepackage{graphicx}
\usepackage{hyperref}
\usepackage{amsmath,amssymb,amsfonts}
\usepackage{graphicx}
\usepackage{textcomp}
\usepackage{xcolor}
\usepackage{cite}
\usepackage{adjustbox}
\usepackage{color, colortbl}
\usepackage{calc}
\usepackage{balance}
\usepackage{subfigure}
\usepackage{xcolor}

\usepackage{amsmath}
\usepackage{pgfplots}

\usepackage{changepage} 

\pgfplotsset{compat=1.10}
\usepgfplotslibrary{fillbetween}

\def\BibTeX{{\rm B\kern-.05em{\sc i\kern-.025em b}\kern-.08emT\kern-.1667em\lower.7ex\hbox{E}\kern-.125emX}}

\usepackage{array}    
\usepackage{booktabs}
\usepackage[skins]{tcolorbox}
\usepackage{tabularx}
\usepackage{multirow}
\usepackage[flushleft]{threeparttable}
\usepackage{amssymb}
\usepackage{enumitem}
\usepackage{rotate}
\usepackage{pgf}
\usepackage{graphicx}
\usepackage{colortbl}
\usepackage{arydshln}
\usepackage{dblfloatfix} 
\usepackage{times}
\usepackage{rotating}
\usepackage{makecell} 
\usepackage{tabularx}
\usepackage{balance}
\usepackage{booktabs}
\usepackage{wrapfig}
\usepackage{tikz}
\usetikzlibrary{angles}
\usepackage{makecell}
\usepackage{tabu}
\newcommand{\bi}{\begin{itemize}}
\newcommand{\ei}{\end{itemize}}
\newcommand{\be}{\begin{enumerate}}
\newcommand{\ee}{\end{enumerate}}

\newcommand{\tbl}[1]{Table~\ref{tbl:#1}}
\newcommand{\eq}[1]{Equation~\ref{eq:#1}}
\newcommand{\tion}[1]{\S\ref{tion:#1}}
\usepackage{amsmath}
\usepackage{subfigure}

\usepackage{arydshln}
\usepackage[hyphenbreaks]{breakurl}
\usepackage{url}

\usepackage{amssymb}
\usepackage{pifont}
\newcommand{\cmark}{\ding{51}}
\newcommand{\xmark}{\ding{55}}

\usepackage{tikz}
\usetikzlibrary{arrows.meta}
\tikzset{%
  >={Latex[width=2mm,length=2mm]},
            base/.style = {rectangle, rounded corners, draw=black,
                           minimum width=2.5cm, minimum height=1cm,
                           text centered, font=\sffamily},
  activityStarts/.style = {base, fill=blue!30},
       startstop/.style = {base, fill=red!30},
    activityRuns/.style = {base, fill=green!30},
         process/.style = {base, minimum width=2.5cm, fill=orange!15,
                           font=\ttfamily},
}

\makeatletter
\def\BState{\State\hskip-\ALG@thistlm}
\makeatother

\newcommand\MyBox[2]{
  \fbox{\lower0.75cm
    \vbox to 1.7cm{\vfil
      \hbox to 1.7cm{\hfil\parbox{1.4cm}{#1\\#2}\hfil}
      \vfil}%
  }%
}

\usepackage[final]{listings}
\usepackage{algorithm}
\usepackage[noend]{algpseudocode}

\newcommand{\crule}[3][darkgray]{\textcolor{#1}{\rule{#2}{#3}}}

\newcommand{\dbox}[1] { \crule[black!#1]{0.67cm}{0.4cm} 
\hspace{-0.51cm}\scalebox{1}[1.0]{{\textcolor{black}{{\bf $^{#1}$}}}\hspace{0.1mm}}}

\usepackage{threeparttable}

\usepackage[framemethod=tikz]{mdframed}
\usetikzlibrary{shadows}
\usepackage{graphics}
\newmdenv[
tikzsetting= {fill=gray!10},
linewidth=1pt,
roundcorner=2pt, 
shadow=false
]{myshadowbox}

\newcommand{\responseChange}[1]{\textcolor{black}{#1}}
\newcommand{\responseChangeTwo}[1]{\textcolor{black}{#1}}

\usepackage[sort&compress,numbers]{natbib}

\newcommand{\IT}{SWIFT}
 
\begin{document}

\title{How to Better Distinguish Security Bug Reports (using Dual Hyperparameter Optimization)}

\author{Rui Shu         \and
        Tianpei Xia     \and
        Jianfeng Chen   \and
        Laurie Williams \and
        Tim Menzies
}

\institute{Rui Shu, Tianpei Xia, Jianfeng Chen, Laurie Williams,
Tim Menzies\at
              Department of Computer Science, North Carolina State University, Raleigh, NC, USA \\
              Email: rshu@ncsu.edu, txia4@ncsu.edu, jchen37@ncsu.edu, lawilli3@ncsu.edu, timm@ieee.org 
}
\date{Received: date / Accepted: date}

\maketitle

\begin{abstract}

\textbf{Background:} In order that the general public is not vulnerable to hackers, security bug reports need to be handled by small groups of engineers before being widely discussed. But learning how to distinguish the security bug reports from other bug reports is challenging since they may occur rarely. Data mining methods that can find such scarce targets require extensive optimization effort.

\textbf{Goal:} The goal of this research is to aid practitioners as they struggle to optimize methods that try to distinguish between rare security bug reports and other bug reports.

\textbf{Method:} Our proposed method, called {\IT}, is a {\em dual optimizer} that optimizes {\em both} learner and pre-processor options. Since this is a large space of options, {\IT} uses a technique called \textit{$\epsilon$-dominance} that learns how to avoid operations that do not significantly improve performance. 

\textbf{Result:} When compared to recent state-of-the-art results (from FARSEC which is published in TSE'18), we find that \responseChangeTwo{the} {\IT}'s dual optimization of both pre-processor and learner is more useful than optimizing each of them individually. For example, in a study of security bug reports from the Chromium dataset, the median recalls of FARSEC and {\IT} were 15.7\% and 77.4\%, respectively. For another example, in experiments with data from the Ambari project, the median recalls improved from 21.5\% to 85.7\% (FARSEC to SWIFT). 

\textbf{Conclusion:} Overall, our approach can quickly optimize models that achieve better recalls than the prior state-of-the-art. These increases in recall are associated with moderate increases in false positive rates (from 8\% to 24\%, median). For future work, these results suggest that dual optimization is both practical and useful.

\keywords{Hyperparameter Optimization \and Data Pre-processing \and Security Bug Report}
\end{abstract}

\section{Introduction}\label{sec:intro}

Security bug detection is a pressing current concern. A report from NIST comments that ``Current systems perform increasingly vital tasks and are widely known to possess vulnerabilities''~\cite{black2016dramatically} (and by ``vulnerability'', they mean a weakness in the computational logic (e.g., code) found in software and some hardware components (e.g., firmware) that, when exploited, results in a negative impact on confidentiality, integrity, or availability~\cite{lewiscommon}). Daily, news reports reveal increasingly sophisticated security breaches. As seen in those reports, a single vulnerability can have devastating effects. For example, a data breach of Equifax caused the personal information of as many as 143 million Americans -- or nearly half the country -- to be compromised~\cite{Equifax}. The WannaCry ransomware attack~\cite{WannaCry} crippled British medical emergency rooms, delaying medical procedures for many patients. 

\begin{figure*}[!htbp]
\centering
\includegraphics[width=10cm]{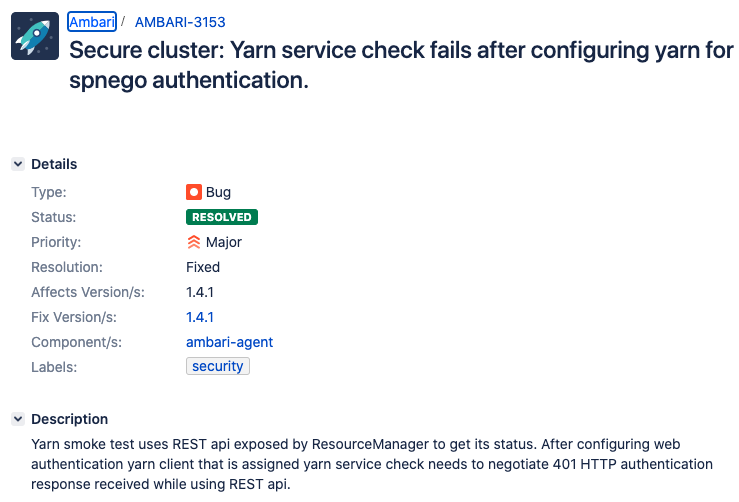}
\caption{An example of security bug report from the Apache Ambari project mislabelled as non-security bug report from~\cite{peters2018text}.}
\label{fig:mislabelledReport}
\end{figure*}

Developers capture and document software bugs and issues into bug reports which are submitted to bug tracking systems. For example, the Mozilla bug database maintains more than 670,000 bug reports with 135 new bug reports added each day~\cite{chen2013r2fix}. Submitted bug reports are explicitly labeled as a security bug report (SBR) or non-security bug report (NSBR). Within such bug tracking systems, Peters et al.~\cite{peters2018text} warn that it is crucial to correctly identify security bug reports and distinguish them from other non-security bug reports. They note that software vendors ask that \responseChange{security bug reports} should be reported directly and privately to their own engineers. These engineers then assess the bug reports and, when necessary, offer a security patch. The security bug, and its associated patch, can then be documented and disclosed via public bug tracking systems. This approach maximizes the probability that a patch is widely available before hackers exploit a vulnerability. \responseChangeTwo{However, due to the lack of security expertise knowledge, bug reporters sometimes mislabel security bug reports as non-security bug reports~\cite{gegick2010identifying}. There are cases when they are not sure when their bug is a non-security bug (which can be safely disclosed) or when that bug is a security bug (that needs to be handled more discretely). For example, Figure~\ref{fig:mislabelledReport} demonstrates a security bug report from the Apache Ambari project, which is mislabelled as non-security bug report. It is a labor intensive process and thus impractical for security practitioners to identify mislabelled security bug reports within a large set of thousands of other non-security bug reports.}

\responseChange{The problem that researchers need to address is how to distinguish security bug reports properly. To tackle this problem, researchers have adopted various techniques. One technique is to apply text mining to the security bug reports~\cite{gegick2010identifying,goseva2018identification,xia2014automated,xia2016automated}. The main idea here is to find some combination of  relevant keywords in the bug reports (as well as features such as word frequency) which are then combined together into classification models. But learning such models is a challenging task since the ratio of security bug reports to other kinds of bug reports may be very low. For example, data sets from~\cite{peters2018text} show among the 45,940 bug reports, only 0.8\% are security bug reports. Various methods exist for mining such rarefied data -- but those methods require extensive optimization effort before they work well on a particular data set. Peters et al. proposed FARSEC~\cite{peters2018text}, a text mining method that used irrelevancy pruning (i.e., filtering). In their approach, developers first identified security related words. Next, they pruned away the irrelevant bug reports (where ``irrelevant'' means ``does not have those security-related keywords''). FARSEC was evaluated using bug reports from one Chromium project and four Apache projects.}

The conjecture of this paper is that this text mining-based method for security bug reports (e.g. as done with FARSEC) can be further enhanced. For example, FARSEC applied its data miners using their default ``off-the-shelf'' configurations. Recently it has been shown that {\em hyperparameter optimization} (which automatically learns the ``magic'' control parameters of an algorithm) can result in better learners that outperform the learners with ``off-the-shelf'' configurations~\cite{agrawal2018wrong,agrawal2018better,fu2016tuning,herodotou2011starfish,tantithamthavorn2016automated,van2017automatic,menzies2019bad}. To the best of our knowledge, this paper is the first attempt to apply hyperparameter optimization to learn models that better  distinguish security bug reports. To that end, we separate and apply three different kinds of optimization strategies:
\be
\item {\em Learner} hyperparameter optimization to adjust the parameters of the data miner; e.g., how many trees to use in random forest, or what values to use in the kernel of Support Vector Machine (SVM).
\item {\em Pre-processor} hyperparameter optimization to adjust any adjustment to the training data, prior to learning; e.g., to learn how to control outlier removal or, how to handle the class imbalance problem.
\item {\em Dual} hyperparameter optimization that combines 1 and 2. 
\ee

\responseChange{Standard practice in the search-based SE literature explores just learner or pre-processor options, but seldom both. There are good reasons for this -- the space of hyperparameters is exponential on the number of optimization options. Hence optimizing {\em both} the learner {\em and} pre-processor options is an exponentially slow process. Nevertheless, this paper shows that if dual optimization can terminate, then it is a useful method. For example, for distinguishing security bug reports, dual optimization performs better than just optimizing learner or pre-processor options individually. This paper succeeds at dual optimization, despite its exponential nature, uses a technique called \textit{$\epsilon$-dominance} to ignore operations that do not significantly improve the performance. We call this method SWIFT in our work.}

\responseChange{In order to demonstrate the efficiency of dual optimization (i.e., {\IT}), we made comparison experiments with the baseline approach (i.e., FARSEC) as well as  state-of-the-art individual optimization methods (i.e., optimizing learners or optimizing pre-processors with the differential evolutionary algorithm). To make that demonstration, we apply dual hyperparameter optimization to the options of \tbl{processorandlearner}. We make no claim that this is the entire set of possible options. Rather, we just say that (a)~any reader of the recent SE data mining literature might have seen many of these; (b)~that reader might be tempted to try optimizing the \tbl{processorandlearner} options; (c)~when we optimize these options in our method, we found that our models were better than the prior state-of-the-art~\cite{peters2018text}.}

\begin{table}[!t]
\centering
\begin{threeparttable}
\caption{List of pre-processors and learners explored in this study. Standard practice in previous literature is to optimize none or just one of these two groups~\cite{bennin2019relative,agrawal2018wrong,agrawal2018better,fu2016tuning,tantithamthavorn2018impact}.
Note that a dual optimizer simultaneously explores both learner and pre-processing options.}
\begin{tabular}{c|l|l}
\hline
\rowcolor[HTML]{ECF4FF} 
\textbf{Type} & \multicolumn{1}{c|}{\cellcolor[HTML]{ECF4FF}\textbf{Name}} & \multicolumn{1}{c}{\cellcolor[HTML]{ECF4FF}\textbf{Description}} \\ \hline
 & Normalizer & Normalize samples individually to unit norm. \\ \cline{2-3} 
 & StandardScalar & \begin{tabular}[c]{@{}l@{}}Standardize features by removing the mean and\\ scaling to unit variance.\end{tabular} \\ \cline{2-3} 
 & MinMaxScaler & \begin{tabular}[c]{@{}l@{}}Transforms features by scaling each feature to \\ a given range.\end{tabular} \\ \cline{2-3} 
 & MaxAbsScaler & Scale each feature by its maximum absolute value. \\ \cline{2-3} 
 & RobustScalar & \begin{tabular}[c]{@{}l@{}}Scale features using statistics that are robust to \\ outliers.\end{tabular} \\ \cline{2-3} 
 & KernelCenterer & Center a kernel matrix. \\ \cline{2-3} 
 & QuantileTransformer & Transform features using quantiles information. \\ \cline{2-3} 
 & PowerTransformer & \begin{tabular}[c]{@{}l@{}}Apply a power transform featurewise to make data\\ more Gaussian-like.\end{tabular} \\ \cline{2-3} 
 & Binarizer & \begin{tabular}[c]{@{}l@{}}Binarize data (set feature values to 0 or 1) according\\ to a threshold.\end{tabular} \\ \cline{2-3} 
 & PolynominalFeatures & Generate polynomial and interaction features. \\ \cline{2-3} 
\multirow{-11}{*}{Pre-processor} & SMOTE & Synthetic Minority Over-sampling Technique. \\ \hline
 & Random Forest (RF) & \begin{tabular}[c]{@{}l@{}}Generate conclusions using multiple entropy-based\\ decision trees.\end{tabular} \\ \cline{2-3} 
 & K Nearest Neighbors (KNN) & \begin{tabular}[c]{@{}l@{}}Classify a new instance by finding ``K'' examples of\\ similar instances.\end{tabular} \\ \cline{2-3} 
 & Naive Bayes (NB) & \begin{tabular}[c]{@{}l@{}}Classify a new instance by (a) collecting mean and\\ standard deviations of attributes in old instances of\\ different classes; (b) returning the class whose\\ attributes are statistically most similar to the new\\ instance.\end{tabular} \\ \cline{2-3} 
 & Logistic Regression (LR) & \begin{tabular}[c]{@{}l@{}}Map the output of a regression into 0 $\leq$ $n$ $\leq$ 1;\\ thus enabling using regression for classification.\end{tabular} \\ \cline{2-3} 
\multirow{-5}{*}{Learner} & Multilayer Perceptron (MLP) & \begin{tabular}[c]{@{}l@{}}A deep artificial neural network which is composed of \\ more than one perceptron.\end{tabular} \\ \hline
\end{tabular}
\begin{tablenotes}
\small
\item Note: The listed pre-processors and learners are based on scikit-learn version 0.21.2. SMOTE is implemented independently without using existing scikit-learn library.
\end{tablenotes}
\label{tbl:processorandlearner}
\end{threeparttable}
\end{table}

This study is structured around the following research questions:

\newenvironment{RQ}{\vspace{2mm}\begin{tcolorbox}[enhanced,width=4.6in,size=fbox,
colback=blue!5,drop shadow southwest,sharp corners]}{\end{tcolorbox}}
    
\begin{RQ}
{\bf RQ1.} \responseChange{Can hyperparameter optimization \responseChangeTwo{techniques} improve the performance of models that better distinguish security bug reports from other bug reports?}
\end{RQ}

\responseChange{We find that the dual hyperparameter optimization approach better distinguishes security bug reports from non-security bug reports. Specifically, our new method increases the recall on the security bug reports from 21.5\% to 66.7\% (median values for FARSEC and {\IT}, respectively). This recall increase is associated with moderate false alarm rate increase from 8.0\% to 24.0\% (median values, FARSEC to {\IT}).} 

\begin{RQ}
{\bf RQ2.} When learning how to  \responseChange{distinguish} security bug reports,
is it better to dual optimize the learners \responseChange{and} the data pre-processors?
\end{RQ}

\responseChange{We will show that dual optimization is statistically significantly better in 31/40 data sets with regard to recall results. This is more than twice as many wins as other approaches explored in this paper. In addition, the dual optimization used here is faster (and scales better to more complex problems) than other techniques.}   
 
\begin{RQ}
\responseChange{{\bf RQ3.} Can hyperparameter optimization further improve the performance of ranking security bug reports?}
\end{RQ}

\responseChange{From the ranking evaluation experiment results, we can observe that individual hyperparameter optimization can achieve better ranking score than the best filter treatment from FARSEC for all five projects studied here. In addition, dual optimization is better than individual optimization in this metric across all five projects.}

In summary, the contributions of this paper are:
\bi
\item {\em \responseChange{An improved result on prior state-of-the-art.}} Specifically, to distinguish security bug reports from non-security bug reports, our methods are better than those reported in the previous FARSEC paper from TSE'18.
\item {\em A comment on the value of optimizing (a) data pre-processors or (b) data mining learners.} Specifically, to identify rare events, we show that dual optimization of (a) and (b) does much better than optimizing either, individually.
\item
{\em A demonstration of the practicality of dual optimization.}
\responseChange{As shown below, the overall runtime for dual optimization (i.e., SWIFT) is five minutes for small datasets and 12 minutes for larger datasets such as the Chromium project on average. This is an important result since our pre-experimental concern is that the cross-product of the option space between the (a) data pre-processors and (b) data mining learners would be so large as to preclude dual optimization.}
\ei

The remainder of this paper is organized as follows. We introduce research background and related work in Section~\ref{sec:background}. We then describe the details of our approach in Section~\ref{tion:swift-e}. In Section~\ref{evaluation}, we present our experiment details, including hyperparameter optimization ranges, datasets, experiment rig, and metrics, etc. We answer proposed research questions in section~\ref{results}. We deliver the take-away messages in Section~\ref{discussion} and discuss the threats to validity in Section~\ref{threats} and then conclude in Section~\ref{conclusion}.

\section{Background and Related Work}\label{tion:vuln}
\label{sec:background}
 
\responseChange{Various methods have been applied to address the need for more secure software. This section first discusses how data mining has been applied to this problem, then we introduce the state-of-the-art FARSEC technique, after which we introduce more details of hyperparameter optimization.}

\subsection{Security Bug Reports and Data Mining}\label{tion:bad}
\responseChange{Data mining has recently been widely applied in bug report analysis, such as identification of duplicated bug reports~\cite{sun2011towards,lazar2014improving,hindle2016contextual,deshmukh2017towards}, prediction of the severity or impact of a reported bug~\cite{lamkanfi2010predicting,zhang2015predicting,tian2012information,yang2016automated,yang2017high}, extraction of execution commands and input parameters from performance bug reports~\cite{han2018perflearner}, assignment of the priority labels to bug reports~\cite{tian2015automated}, bug report field reassignment and refinement prediction~\cite{xia2016automated} and identify vulnerabilities from commit message and bug reports~\cite{zhou2017automated} }.

In particular, a few studies of bug report classification are more relevant to our work. Some of those approaches focus on building bug classification models based on analyzing bug reports with text mining. For example, Zhou et al.~\cite{zhou2016combining} leveraged text mining techniques, analyzed the summary parts of bug reports and fed into machine learners. Xia et al.~\cite{xia2014automated} developed a framework that applied text mining technology on bug reports and trained a model on bug reports with known labels (i.e., configuration or non-configuration). The trained model was used to predict the new bug reports. Popstojanova et al.~\cite{goseva2018identification} used different types of textual feature vectors and focused on applying both supervised and unsupervised algorithms in classifying security and non-security related bug reports. Wijayasekara et al.~\cite{wijayasekara2014vulnerability} extracted textual information by utilizing the textual description of the bug reports. A feature vector was generated through the textual information and then presented to a machine learning classifier.

Some other approaches use a more heuristic way to identify bug reports. For example, Zaman et al.~\cite{zaman2011security} combined keyword searching and statistical sampling to distinguish between performance bugs and security bugs in Firefox bug reports. Gegick et al.~\cite{gegick2010identifying} proposed a technique to identify security bug reports based on keyword mining and performed an empirical study based on an industry bug repository.

\responseChange{ 
While all the above work significantly advanced the state-of-the-art, but results related to data mining on software security issues are often problematic:
\bi
\item
Neuhaus \& Zimmermann ~\cite{neuhaus2007predicting} explored the dependency structure within RedHat Linux to learn vulnerability predictors with precision and recall of 83\% and 65\%. Neuhaus \& Zimmermann~\cite{neuhaus2009beauty} later applied their dependency-based methods to the same code base, but at a much larger scale of granularity (system, not specific applications). Their results were not impressive: precision and recall of 40\% and 20\%, respectively.
\item
Nguyen \& Tran~\cite{nguyen2010predicting}, similarly, applied explored dependency structure. Though not as impressive as Neuhaus and Zimmermann, they achieved precision and recall of 60\% and 61\%. However, their code dependency network analysis is not a general method for building vulnerability predictors.
\item
Scandariato et al.~\cite{scandariato2014predicting} used a text mining approach over the source code for their vulnerability predictors. They report prediction models with precision and recall over 95\%. However, these results were based on a somewhat contentious methodology. The unfiltered alerts of a static code analysis tool were used to label code components as ``vulnerable'' or not. Such static code analysis tools have a notoriously large false positive rate, declaring that many code components are ``vulnerable'' when the vulnerabilities are actually false positives.
\ei}

\subsection{FARSEC: Extending Data Mining for Bug Reports} \label{tion:farsec}

\responseChange{The previous section reported certain problems with existing
methods where data mining was applied to security related tasks. In the recently proposed FARSEC~\cite{peters2018text} research, Peters et al. reported more success after focusing on a particular problem within the security domain.}

FARSEC is a technique that adds an irrelevancy pruning step to data mining in building security bug prediction models. \tbl{farsecFilter} lists the filters explored in the FARSEC research. The purpose of filtering in FARSEC is to remove non-security bug reports with security related keywords. To achieve this goal, FARSEC applied an algorithm that firstly calculated the probability of the keywords appearing in security bug report and non-security bug report, and then calculated the score of the keywords.

Inspired by previous works~\cite{graham2004hackers,jalali2008optimizing}, several tricks were also introduced in FARSEC to reduce false positives. For example, FARSEC built the {\em farsectwo} filter by multiplying the frequency of non-security bug reports by two, aiming to achieve a good bias. The {\em farsecsq} filter was created by squaring the numerator of the support function to improve heuristic ranking of low frequency evidence.  

\begin{table*}[!htbp]
\caption {Different filters used in FARSEC.}
\centering
\begin{tabular}{l|l}
\hline
\rowcolor[HTML]{ECF4FF} 
\multicolumn{1}{c|}{\cellcolor[HTML]{ECF4FF}\textbf{Filter}} & \multicolumn{1}{c}{\cellcolor[HTML]{ECF4FF}\textbf{Description}} \\ \hline
farsecsq & \begin{tabular}[c]{@{}l@{}}Apply the Jalali et al.~\cite{jalali2008optimizing} support function\\ to the frequency of words found in SBRs\end{tabular} \\ \hline
farsectwo & \begin{tabular}[c]{@{}l@{}}Apply the Graham version~\cite{graham2004hackers} of multiplying \\ the frequency by two.\end{tabular} \\ \hline
farsec & Apply no support function. \\ \hline
clni & Apply CLNI filter to non-filtered data. \\ \hline
clnifarsec & Apply CLNI filter to farsec filtered data. \\ \hline
clnifarsecsq & Apply CLNI filter to farsecsq filtered data. \\ \hline
clnifarsectwo & Apply CLNI filter to farsectwo filtered data. \\ \hline
\end{tabular}
\label{tbl:farsecFilter}
\end{table*}

In addition, FARSEC also tested a noise detection algorithm called CLNI (Closet List Noise Identification)~\cite{kim2011dealing}. Specifically, CLNI works as follows: During each iteration, for each instance $i$, a list of closest instances are calculated and sorted according to Euclidean Distance to instance $i$. The percentage of top $N$ instances with different class values is recorded. If percentage value is larger or equal to a threshold, then instance $i$ is highly probable to be a noisy instance and thus included to noise set $S$. This process is repeated until two noise sets $S_{i}$ and $S_{i-1}$ have the similarity over $\epsilon$ \responseChange{(e.g., $\epsilon$ is 0.99)}. A threshold score (e.g., $0.75$) is set to remove any non-buggy reports above the score.

\responseChange{One of the common issues with imbalanced data prediction is the large number of false positives in the prediction results. This matters because it means potentially extra effort is required from developers to check those false positives. FARSEC tries to address this problem by generating a list of ranked bug reports. This method takes two steps. In the first step, for a filter $f$, the ranked prediction results are selected from non-filtered data or data with filters other than $f$ which has less number of predicted security bug reports than filter $f$. If the first step does not apply, the chronological order is used in step two. As a result, the predicted security bug reports are close to the top of the list than non-security bug reports.}

\subsection{Hyperparameter Optimization for Learner and Pre-Processor Options}\label{tion:hpo}

\responseChange{One data mining approach not fully explored by FARSEC
(or much of other works reviewed above) is hyperparameter optimization,
i.e. the process of searching the most optimal hyperparameters in data mining learners~\cite{biedenkapp2018hyperparameter}. In machine learning, hyperparameters reflect policies within a model. For example:}
\responseChange{
\bi
\item
For random forest, a hyperparameter could be the number of trees in the forest.
\item
For nearest neighbor algorithm, a hyperparameter could be the number of $k$ nearest neighbors used for classification~\cite{keller1985fuzzy}.
\item
For text mining, a hyperparameter might control how many words are selected via term weighting.
\ei
}

\begin{table*}[!htbp]
\centering
\caption {List of hyperparameters optimized in different learners and pre-processors. The brief description of each learner and pre-processor can be found in Table~\ref{tbl:processorandlearner}.}
\begin{tabular}{c|l|l|c|c}
\hline
\rowcolor[HTML]{ECF4FF} 
\textbf{Type} & \multicolumn{1}{c|}{\cellcolor[HTML]{ECF4FF}\textbf{Name}} & \multicolumn{1}{c|}{\cellcolor[HTML]{ECF4FF}\textbf{Parameters}} & \textbf{Default} & \textbf{Tuning Range} \\ \hline
 &  & n\_estimators & 10 & {[}10, 150{]} \\  
 &  & min\_samples\_leaf & 1 & {[}1, 20{]} \\  
 &  & min\_samples\_split & 2 & {[}2, 20{]}  \\  
 &  & max\_leaf\_nodes & None & {[}2, 50{]} \\  
 &  & max\_features & auto & {[}0.01, 1{]}  \\  
 & \multirow{-6}{*}{Random Forest} & max\_depth & None & {[}1, 10{]} \\ \cline{2-5} 
 &  & C & 1.0 & {[}1.0, 10.0{]} \\  
 & \multirow{-2}{*}{Logistic Regression} & max\_iter & 100 & {[}50, 200{]} \\ \cline{2-5} 
 &  & alpha & 0.0001 & {[}0.0001, 0.001{]} \\  
 &  & learning\_rate\_init & 0.001 & {[}0.001, 0.01{]} \\  
 &  & power\_t & 0.5 & {[}0.1, 1{]}  \\  
 &  & max\_iter & 200 & {[}50, 300{]}  \\  
 &  & momentum & 0.9 & {[}0.1, 1{]}  \\  
 & \multirow{-6}{*}{Multilayer Perceptron} & n\_iter\_no\_change & 10 & {[}1, 100{]}  \\ \cline{2-5} 
 &  & leaf\_size & 30 & {[}10, 100{]}  \\  
 & \multirow{-2}{*}{K Nearest Neighbor} & n\_neighbors & 5 & {[}1, 10{]}  \\ \cline{2-5} 
\multirow{-18}{*}{Learner} & Naive Bayes & var\_smoothing & 1e-9 & {[}0.0, 1.0{]}  \\ \hline
 &  & k & 5 & {[}1, 20{]} \\  
 &  & m & 50\% & {[}50, 400{]} \\  
 & \multirow{-3}{*}{SMOTE} & r & 2 & {[}1, 6{]}  \\ \cline{2-5} 
 &  & norm & l2 & {[}l1, l2, max{]} \\  
 & \multirow{-2}{*}{Normalizer} & copy & True & {[}True, False{]} \\ \cline{2-5} 
 &  & copy & True & {[}True, False{]}\\  
 &  & with\_mean & True & {[}True, False{]}  \\  
 & \multirow{-3}{*}{StandardScaler} & with\_std & True & {[}True, False{]}  \\ \cline{2-5} 
 &  & copy & True & {[}True, False{]} \\  
 &  & min & 0 & {[}-5, 0{]}  \\  
 & \multirow{-3}{*}{MinMaxScaler} & max & 1 & {[}1, 5{]}  \\ \cline{2-5} 
 & MaxAbsScaler & copy & True & {[}True, False{]}  \\ \cline{2-5} 
 &  & with\_centering & True & {[}True, False{]}  \\  
 &  & with\_scaling & True & {[}True, False{]}  \\  
 &  & q\_min & 25.0 & {[}10, 40{]}  \\  
 &  & q\_max & 75.0 & {[}60, 90{]}  \\  
 & \multirow{-5}{*}{RobustScaler} & copy & True & {[}True, False{]} \\ \cline{2-5} 
 &  & n\_quantiles & 1000 & {[}10, 2000{]}  \\  
 &  & output\_distribution & uniform & {[}uniform, normal{]}  \\  
 &  & ignore\_implicit\_zeros & False & {[}True, False{]}  \\  
 &  & subsample & 1e5 & {[}100, 150000{]} \\  
 & \multirow{-5}{*}{QuantileTransformer} & copy & True & {[}True, False{]}  \\ \cline{2-5} 
 &  & method & yeo-johnson & \begin{tabular}[c]{@{}c@{}}{[}yeo-johnson, \\ box-cox{]}\end{tabular} \\  
 &  & standardize & True & {[}True, False{]}  \\  
 & \multirow{-3}{*}{PowerTransformer} & copy & True & {[}True, False{]}  \\ \cline{2-5} 
 &  & threshold & 0.0 & {[}0, 10{]}  \\  
 & \multirow{-2}{*}{Binarization} & copy & True & {[}True, False{]}  \\ \cline{2-5} 
 &  & degree & 2 & {[}2, 4{]} \\  
 &  & interaction\_only & False & {[}True, False{]}  \\  
 &  & include\_bias & True & {[}True, False{]} \\  
\multirow{-31}{*}{Pre-processor} & \multirow{-4}{*}{PolynomialFeatures} & order & C & {[}C, F{]} \\ \hline
\end{tabular}
\label{tbl:hyperparameter}
\end{table*}

\responseChange{In this list, the first two are examples of learner hyperparameters while the third one is an example of pre-processor hyperparameter that is selected before the learner executes. Table~\ref{tbl:processorandlearner} lists the learner and pre-processor options we explore in this study. The search space of these parameters is shown in \tbl{hyperparameter}. In those tables, we use the same five machine learning learners as seen in the FARSEC study, i.e., Random Forest (RF), Naive Bayes (NB), Logistic Regression (LR), Multilayer Perceptron (MLP) and K Nearest Neighbor (KNN). They are widely used for software engineering classification problems~\cite{lessmann2008benchmarking}. As for the pre-processors, as mentioned in the introduction section, we do not claim that this is the entire set of possible pre-processors. Rather, we just say that  any reader of the recent SE data mining literature might have seen many of these. Hence, they might be tempted to try them.}

\responseChange{Furthermore, Table~\ref{tbl:previousWork} shows how often these kinds of hyperparameters have been explored in the previous security relevant literature. As seen from the table:
\bi
\item
A minority of papers have explored learner hyperparameter optimization.
\item
Only a handful of them have tried pre-processor hyperparameter optimization.
\item 
We have only found one prior work that tried our {\em dual optimization} approach that explored both pre-processor and learner optimization~\cite{agrawal2019dodge}. However, note that that paper was not in the security domain.
\ei
}

\responseChange{There are good reasons to try and avoid dual optimization -- an exhaustive search through all options is computationally intractable. Given $N$ choices for $P$ learner parameters, the space of possible hyperparameter optimizations in $(N)^P$. Worse still, if the space of options increases to include learners and $N$ choices for $M$ pre-processors (such as those listed in \tbl{processorandlearner}), then the search space is now $(N)^{P+M}$, i.e. exponentially larger.}

\begin{table*}[!htbp]
\centering
\caption{List of previous research studies that address security and software engineering problems. In this list, only one prior publication
optimized both the learner and pre-processor (see the
last line, \colorbox{black!15}{highlighted in gray}) and that paper did not explore the security domain.
This list of papers was found either from the above literature review or from Google Scholar using the search query, e.g., {\em ``((hyperparameter optimization) and (security)) or ((hyperparameter optimization) and (security bug reports)) or (optimization and security) or (optimization and (pre-processors) and security), ((hyperparameter optimization) and (software engineering))''}.
These queries returned more than 5,000 papers which were further pruned. We only used papers in the last ten years (2010-2020) and which had appeared in (a)~top conferences or (b) ~venues listed by Google Scholar as ``top-ranked'' (e.g., see \textcolor{blue}{\href{http://tiny.cc/top20soft_venues}{tiny.cc/top20soft\_venues}}).}
\begin{tabular}{c|c|c|c|c|c}
\hline
\rowcolor[HTML]{ECF4FF} 
\textbf{Reference} & \textbf{Year} & \textbf{Citation} & \textbf{\begin{tabular}[c]{@{}c@{}}Learner\\ Optimization\end{tabular}} & \textbf{\begin{tabular}[c]{@{}c@{}}Pre-processor\\ Optimization\end{tabular}} & \textbf{\begin{tabular}[c]{@{}c@{}}Security\\ Related\end{tabular}} \\ \hline
\cite{thornton2013auto} & 2013 & 754 & \cmark & \xmark & \xmark \\ 
\cite{li2017hyperband} & 2017 & 358 & \cmark & \xmark & \xmark \\ 
\cite{lamkanfi2010predicting} & 2010 & 285 & \xmark & \xmark & \cmark \\
\cite{sun2011towards} & 2011 & 264 & \xmark & \xmark & \xmark \\
\cite{feurer2015initializing} & 2015 & 193 & \cmark & \xmark & \xmark \\ 
\cite{gegick2010identifying} & 2010 & 146 & \xmark & \xmark & \cmark \\
\cite{xia2017boosted} & 2017 & 139 &\cmark & \xmark & \xmark \\ 
\cite{tian2012information} & 2012 & 133 & \cmark & \xmark & \cmark \\
\cite{fu2016tuning} & 2016 & 100 & \cmark & \xmark & \xmark \\
\cite{tian2015automated} & 2015 & 64 & \xmark & \xmark & \cmark \\
\cite{wang2018leveraging} & 2018 & 60 & \cmark & \xmark & \xmark \\ 
\cite{agrawal2018wrong} & 2018 & 59 & \xmark & \cmark & \cmark \\
\cite{lazar2014improving} & 2014 & 54 & \xmark & \xmark & \xmark \\
\cite{agrawal2018better} & 2018 & 49 & \xmark & \cmark & \xmark \\
\cite{xia2014automated} & 2014 & 44 & \xmark & \xmark & \xmark \\
\cite{tantithamthavorn2018impact} & 2018 & 34 & \xmark & \cmark & \xmark \\
\cite{hindle2016contextual} & 2016 & 29 & \xmark & \xmark & \xmark \\
\cite{nair2018finding} & 2018 & 29 & \cmark & \xmark & \xmark \\
\cite{zhang2015predicting} & 2015 & 28 & \xmark & \xmark & \cmark \\
\cite{wijayasekara2014vulnerability} & 2014 & 26 & \xmark & \xmark & \cmark \\
\cite{yang2017high} & 2017 & 23 & \xmark & \xmark & \cmark \\
\cite{chan2013continuous} & 2013 & 20 & \cmark & \xmark & \xmark \\ 
\cite{di2018genetic} & 2018 & 20 & \cmark & \xmark & \xmark \\ 
\cite{deshmukh2017towards} & 2017 & 18 & \cmark & \xmark & \xmark \\
\cite{xia2016automated} & 2016 & 17 & \xmark & \xmark & \xmark \\
\cite{osman2017hyperparameter} & 2017 & 14 & \cmark & \xmark & \cmark \\
\cite{menzies2018500+} & 2018 & 16 & \cmark & \xmark & \xmark \\
\cite{yang2016automated} & 2016 & 11 & \xmark & \xmark & \cmark \\
\cite{goseva2018identification} & 2018 & 9 & \xmark & \xmark & \cmark \\
\cite{han2018perflearner} & 2018 & 6 & \xmark & \xmark & \xmark \\
\rowcolor{black!15} \cite{agrawal2019dodge} & 2019 & 4 & \cmark & \cmark & \xmark \\ \hline
\end{tabular}
\label{tbl:previousWork}
\end{table*}

It is neither useful nor practical to explore such a large space of options via exhaustive search. For example, \textit{grid search}~\cite{bergstra2011algorithms, tantithamthavorn2016automated} is a ``brute force'' hyperparameter optimizer that wraps a learner into for-loops that walk through a wide range of all learner's control parameters. Simple to implement, it has many drawbacks. Firstly, even this brute force approach does not sample all the options since its for-loops jump over numeric ranges using some increment value. This means that grid search can actually skip over the important optimizations. Secondly, it suffers from the ``curse of dimensionality''. That is, after just a  handful of options, grid search can miss important optimizations. Thirdly, and worse still, much CPU resources can be wasted during grid search since experience has shown that only a few ranges within a few optimization parameters really matter~\cite{bergstra2012random}.

An alternative to grid search is the \textit{random search}~\cite{bergstra2012random} that stochastically samples the search space and evaluates sets from a specified probability distribution. Evolutionary algorithms are a variant of random search that runs in ``generations'' where each new generation is seeded from the best examples selected from the last generation~\cite{goldberg2006genetic}. Simulated annealing is a special form of evolutionary algorithms where the population size is one~\cite{kirkpatrick1983optimization,Menzies:2007a}. 

\textit{Genetic algorithms (GA)} is another form of random search where the population size is greater than one, and new mutants are created by crossing over parts of the better members of the current population~\cite{goldberg2006genetic,Panichella:2013}.
Note one feature of genetic algorithms is that, their mutation operator never changes during the execution of the GA. That is, GAs have no facility for using experience from the domain to define better mutators.

Another kind of random search, that does use domain experience to define better mutators, is {\em differential evolution} (DE)~\cite{storn1997differential}. In differential evolution algorithm, the size of a mutation is selected from a pool of previous cache of ``superior'' mutations; i.e. mutants that are known to be better than other mutants. That is, as differential evolution algorithm learns more and more about what mutants are superior, it is also learning how better to mutate old individuals into better ones. \responseChange{ There are four major steps in differential evolution algorithm -- {\em initialization}, \textit{mutation}, \textit{crossover}, and {\em selection}:
\bi
\item
The {\em initialization} step creates a population of individuals, while each individual is an instance of the parameters generated randomly within given bounds.
\item
In the {\em mutation} step, for each individual $p_i$ in the population, three other individuals $a, b, c$ (not the current one) are randomly selected. A mutant individual is created by combining these three selected individuals. The difference is then computed between two individuals and added to the rest individual after multiplying a mutation factor to the difference, i.e., \mbox{$y_k = a_k + f \times (b_k - c_k)$}. The mutation factor $f$ is a positive number that controls the amplification difference between two individuals.
\item
At some crossover probability $\mathit{cf}$, the mutant attribute is then added to a vector that is the new mutant in the $crossover$ step.
\item
Finally, during the {\em selection} step, differential evolution algorithm decides if the mutant generated from $a,b,c$ is better than $p_i$. If so, the mutant replaces $p_i$ and the algorithm moves on to some other member of the population $p_j$.
\item
All the above steps have to be repeated again for the remaining individuals $p_j$, which completes the first iteration of the algorithm. After this process, some of the original individuals of the population will be replaced by better ones. That is, all subsequent mutants will be built from the ``superior'' examples cached in the population. 
\ei 
}

\responseChange{As to the control parameters of the differential evolution algorithm, using advice from the differential evolution algorithm user group (see \href{http://tiny.cc/how2de}{tiny.cc/how2de}), we set $\{\mathit{np},f,cr\}=\{10k,0.8,0.9\}$, where $k$ is the number of parameters to optimize, and $np$ is the size of whole population. Note that we set the number of iteration $\{g\}$ to $3$, $10$, which are denoted as DE$3$ and DE$10$ respectively. A small number (i.e., $3$) is used to test the effects of a CPU-light effort estimator. A larger number (i.e., $10$) is selected to check if anything is lost by restricting the inference to small iterations.}

In the software engineering literature, differential evolution algorithm has been seen to outperform other methods such as (a)~particle swarm optimization~\cite{vesterstrom2004comparative}; (b)~the grid search used by Tantithamthavorn et al.~\cite{tantithamthavorn2016automated} to optimize their defect predictors; or (c)~the genetic algorithm used by Panichella et al.~\cite{Panichella:2013} to optimize a text miner. Also, the differential evolution algorithm has been proven useful in prior software engineering optimization studies~\cite{fu2016tuning}.

\section{{\IT}: the Dual Optimization Approach}
\label{tion:swift-e}

\responseChangeTwo{Recent studies show substantial interest in automated hyperparameter optimization on complex and computational expensive machine learning models with many hyperparameters. By tailoring the models to the problems at hand, hyperparameter optimization improves the model performance and even leads to new state-of-the-art results.}

\responseChangeTwo{Apart from machine learning models, data pre-processing techniques are often involved in practical machine learning pipeline. Real-world data is often inconsistent, lacking in certain behaviors of trends, or even contains many errors. Data pre-processing transforms the raw data into a more useful and efficient shape. Similar to model optimization, pre-processing optimization also shows increasing interest~\cite{agrawal2018better}.}

\responseChangeTwo{While each individual optimization problem already experiences computational complexity, for example, Table~\ref{tbl:processorandlearner} and Table~\ref{tbl:hyperparameter} demonstrate a list of machine learning learners and data pre-processing techniques, as well as their hyperparameter options. Even this partial list includes thousands of configuration options. The cost of running an optimizer through these options would be quite expensive, requiring days to weeks of CPU resources~\cite{tantithamthavorn2016automated}~\cite{tantithamthavorn2018impact}. A combination of the above two optimization problems (i.e., dual optimization) faces even more challenges.}

\responseChangeTwo{A ``simpler'' optimizer is required to tackle the dual optimization challenge. This ideal optimizer should be able to achieve better performance than each individual optimizer and the computational complexity would not increase.}

\responseChangeTwo{In 2005, Deb et al.~\cite{deb2005evaluating} proposed an idea named \textit{$\epsilon$-dominance} that partitions the output space of an optimizer into $\epsilon$-sized grids. The principle of this idea is that if there exists some $\epsilon$ value below which it is useless or impossible to distinguish the results, then it superfluous to explore anything less than $\epsilon$. Specifically, consider the  bug reports classification task discussed in this paper, if the performances of two learners (or a learner with various hyperparameters) differ in less than some $\epsilon$ value, then we cannot statistically distinguish them. For the learners which do not significantly improve the performance, we can further reduce the attention on them.}

\responseChangeTwo{Inspired by the idea of $\epsilon$-dominance, we propose a method named SWIFT to address the dual optimization problem. From a high level, SWIFT is essentially a tabu search; i.e., if some settings resulted in some performance within $\epsilon$ of any older result, then SWIFT marked that option as ``to be avoided''. SWIFT applies ``item ranking'' in seeking optimal learner and pre-processor, and further refines their option ranges. SWIFT returned the best setting seen during the following three stage process:}

\bi
\item {\em Initialization}: all option items $i$ are assigned equal weightings.
\item The {\em item ranking} stage reweights items $i$ in column 2 of Table~\ref{tbl:hyperparameter}; e.g. terms like ``Random Forest'' or ``RobustScaler''.
\item The {\em numeric refinement} stage adjusts the tuning ranges of the last column in Table~\ref{tbl:hyperparameter}.
\ei

In summary, what is happening here is that item selection handles the ``big picture'' decisions about what pre-processor or learner to use while   numeric refinement focuses on smaller details about numeric ranges.

More specifically, the algorithm runs as follows:
\bi
\item {\em Initialization}: Assign weights $w_i=0$ to all items $i$ in column 2 of Table~\ref{tbl:hyperparameter}.
\item {\em Item ranking}: $N_1$ times, we make a random selection of a learner and pre-processor from column 2, favoring those items with higher weights. For the selected items, we select a value at random from the  ``Tuning Range''s of the last column of Table~\ref{tbl:hyperparameter}. Using that selection, we build a model and evaluate it on test data. If we obtain a model whose performance is more/less than $\epsilon$ of any prior results, then we  add/subtract (respectively) 1.0  from $w_i$.
\item {\em Numeric refinement}: $N_2$ times, we refine the numeric tuning ranges $(\mathit{lo}, \mathit{hi})$ seen in the last column of Table~\ref{tbl:hyperparameter}. In this step, the item ranking continues. But now, if ever some numeric tuning value $\mathit{lo} \le b \le \mathit{hi}$ produces a better model, then we adjust that range, as follows.
Whichever of $x \in (${\em lo, hi}) that is the furthest from $b$ is moved to $(b+x)/2$.
\ei

\responseChangeTwo{(Aside: It should be pointed out that SWIFT is not a multi-objective optimization problem. We choose g-measure as our optimization goal (i.e., the aim to increase). G-measure is the harmonic mean of recall and the complement of false alarms. More description of this metric and the reason of the choice are further discussed in Section~\ref{sec:metrics}.)}

\responseChangeTwo{Agrawal et al.~\cite{agrawal2019dodge} have successfully applied $\epsilon$-dominance to some SE tasks such as software defect prediction and SE text mining, and they proposed the approach named DODGE. For the cases studied by DODGE, that approach was able to explore a large space of hyperparameter options, while at the same time generated models that performed as well or better than the prior state-of-the-art in defect prediction and SE text mining~\cite{agrawal2019dodge}. SWIFT is an improved version of DODGE since we found that DODGE cannot be directly applied to our bug report data without any modification effort. There are several reasons for this after investigation.} 

\responseChange{Firstly, DODGE guided its optimization using metrics that were alien to this domain. For example, the ``Popt20'' goal used in the original DODGE studied by Agrawal et al.~\cite{agrawal2019dodge} optimizes for an economic concern not explored by Peters et al. in the FARSEC study. Popt20 is relevant to general SE tasks, but not for security-related domains. Specifically, we want to find as many of the security bug reports as possible, even if that means developers have to spend some time exploring a few more false positives. Accordingly, we swapped out Popt20 in favor of the ``g-measure'' as defined in Section~\ref{evaluation}.}

\responseChange{Second, once we changed evaluation goals, another concern became apparent. We found that the distribution of the $w_i$ weights was far more  skewed in the security bug report data than in the other kinds of software engineering tasks studied by Agrawal et al. This skewed data meant that, usually, there was only one good learner and one good data pre-processor for the security data sets. We conjecture that this is so since we require specific biases to find the target concept of something so particular as a security bug report. For the original version of DODGE, such skewed $w_i$ weights are a problem since, as mentioned above, item ranking continues during the numeric refinement stage.}

\responseChange{{\IT} is specifically designed for our security data. {\IT} is designed to make better use of the $w_i$ skews. After item ranking, {\IT} only takes the best learner and data pre-processor forward into numeric refinement. While the above two changes were only a  small coding change to the original DODGE, their effects were profound.}

\section{Experiment}
\label{evaluation}

\subsection{Hyperparameter Optimization Ranges}\label{tion:range}

This paper compares {\IT} against the differential evolution algorithm (described in Section~\ref{sec:background}) since recent papers at ICSE~\cite{agrawal2018better} the IST journal~\cite{agrawal2018wrong} reported that the differential evolution algorithm can find large improvement in learner performance for SE data. \tbl{DEpara} lists the control settings for the differential evolution algorithm used in this paper
(that table was generated by combining the advice at the end of \tion{hpo} with Table~\ref{tbl:hyperparameter}). For {\IT}, we used the
settings recommended by Agrawal et al.~\cite{agrawal2019dodge}. \responseChangeTwo{Note that proving the optimum of our solution is not the goal of this paper. In fact, like Wolpert~\cite{wolpert1997no},
we doubt if there is any  ``best'' optimizer that works for
all data
(for more on that, see the ``No Free Lunch'' theorem discussion~\cite{wolpert1997no} in search and optimization). Therefore, this paper is not searching for the ``best'' result, but rather it is searching for ``better''   than the prior state-of-the-art.}

\begin{table*}[!htbp]
\caption {List of parameters in differential evolution (DE) algorithm for different learners and pre-processor.}
\centering
\begin{threeparttable}
\begin{tabular}{l|c|c|c|c}
\hline
\rowcolor[HTML]{ECF4FF} 
\multicolumn{1}{c|}{\cellcolor[HTML]{ECF4FF}} & \multicolumn{4}{c}{\cellcolor[HTML]{ECF4FF}\textbf{DE Parameter}} \\ \cline{2-5} 
\rowcolor[HTML]{ECF4FF} 
\multicolumn{1}{c|}{\multirow{-2}{*}{\cellcolor[HTML]{ECF4FF}\textbf{Learner \& Pre-processor}}} & \textbf{NP} & \textbf{F} & \textbf{CR} & \textbf{ITER} \\ \hline
Random Forest & 60 &  &  &  \\ \cline{1-2}
Logistic Regression & 30 &  &  &  \\ \cline{1-2}
Multilayer Perceptron & 60 &  &  &  \\ \cline{1-2}
K Nearest Neighbor & 20 &  &  &  \\ \cline{1-2}
Naive Bayes & 10 & \multirow{-5}{*}{0.8} & \multirow{-5}{*}{0.9} & \multirow{-5}{*}{3, 10} \\ \hline
SMOTE & 30 & 0.8 & 0.9 & 10 \\ \hline
\end{tabular}
\begin{tablenotes}
      \item * Note: \textbf{NP} is the size of population; \textbf{F} is the parameter controlling the differential weight; \textbf{CR} is the probability threshold; \textbf{ITER} is the number of iterations.
    \end{tablenotes}
  \end{threeparttable}
\label{tbl:DEpara}
\end{table*}

Note that {\IT} and differential evolution algorithm were applied to learners from the scikit-learn toolkit~\cite{pedregosa2011scikit}. Table~\ref{tbl:hyperparameter} lists all the hyperparameters we select for both data mining learners and data pre-processors based on scikit-learn. 

\responseChangeTwo{We choose not to explore other hyperparameter optimizers, for pragmatic reasons. Numerous other studies have shown that the differential evolutionary algorithm (DE) well performed for optimization problems~\cite{menzies2018500+}~\cite{fu2017easy}~\cite{fu2016tuning}~\cite{wang2015back}~\cite{yildizdan2020novel}~\cite{onan2016multiobjective}. If our goal was to claim that DE was somehow the optimal optimizer, we would have to perform a wider range of study of optimizers (i.e more than just DE). However, our goal is not that (and, in fact, there are support theoretical reasons for assuming that no optimizer is ever ``best'' for all data sets~\cite{wolpert1997no}). Rather, our purpose is to provide an improvement on the prior state-of-the-art (the FARSEC paper). As shown below, that can be achieved using DE. While in future work we aim to explore other optimizers, for the purposes of this paper, using DE is enough.}

\begin{table*}
\caption {Imbalanced characteristic of bug report data sets from FARSEC~\cite{peters2018text}.}
\centering
\begin{tabular}{c|l|c|r|c|c|r|c}
\hline
\rowcolor[HTML]{ECF4FF} 
\cellcolor[HTML]{ECF4FF} & \multicolumn{1}{c|}{\cellcolor[HTML]{ECF4FF}} & \multicolumn{3}{c|}{\cellcolor[HTML]{ECF4FF}\textbf{Training}} & \multicolumn{3}{c}{\cellcolor[HTML]{ECF4FF}\textbf{Testing}} \\ \cline{3-8} 
\rowcolor[HTML]{ECF4FF} 
\multirow{-2}{*}{\cellcolor[HTML]{ECF4FF}\textbf{Project}} & \multicolumn{1}{c|}{\multirow{-2}{*}{\cellcolor[HTML]{ECF4FF}\textbf{Filter}}} & \textbf{\#SBRs} & \multicolumn{1}{c|}{\cellcolor[HTML]{ECF4FF}\textbf{\#BRs}} & \textbf{SBRs(\%)} & \textbf{\#SBRs} & \multicolumn{1}{c|}{\cellcolor[HTML]{ECF4FF}\textbf{\#BRs}} & \textbf{SBRs(\%)} \\ \hline
 & train &  & 20,970 & 0.37 &  &  &  \\ 
 & farsecsq &  & 14,219 & 0.54 &  &  &  \\ 
 & farsectwo &  & 20,968 & 0.37 &  &  &  \\ 
 & farsec &  & 20,969 & 0.37 &  &  &  \\ 
 & clni &  & 20,154 & 0.38 &  &  &  \\ 
 & clnifarsecsq &  & 13,705 & 0.56 &  &  &  \\ 
 & clnifarsectwo &  & 20,152 & 0.38 &  &  &  \\ 
\multirow{-8}{*}{Chromium} & clnifarsec & \multirow{-8}{*}{77} & 20,153 & 0.38 & \multirow{-8}{*}{115} & \multirow{-8}{*}{20,970} & \multirow{-8}{*}{0.55} \\ \hline
 & train &  & 500 & 0.80 &  &  &  \\
 & farsecsq &  & 136 & 2.94 &  &  &  \\ 
 & farsectwo &  & 143 & 2.80 &  &  &  \\ 
 & farsec &  & 302 & 1.32 &  &  &  \\ 
 & clni &  & 392 & 1.02 &  &  &  \\ 
 & clnifarsecsq &  & 46 & 8.70 &  &  &  \\ 
 & clnifarsectwo &  & 49 & 8.16 &  &  &  \\ 
\multirow{-8}{*}{Wicket} & clnifarsec & \multirow{-8}{*}{4} & 196 & 2.04 & \multirow{-8}{*}{6} & \multirow{-8}{*}{500} & \multirow{-8}{*}{1.20} \\ \hline
 & train &  & 500 & 4.40 &  &  &  \\ 
 & farsecsq &  & 149 & 14.77 &  &  &  \\ 
 & farsectwo &  & 260 & 8.46 &  &  &  \\ 
 & farsec &  & 462 & 4.76 &  &  &  \\ 
 & clni &  & 409 & 5.38 &  &  &  \\ 
 & clnifarsecsq &  & 76 & 28.95 &  &  &  \\ 
 & clnifarsectwo &  & 181 & 12.15 &  &  &  \\ 
\multirow{-8}{*}{Ambari} & clnifarsec & \multirow{-8}{*}{22} & 376 & 5.85 & \multirow{-8}{*}{7} & \multirow{-8}{*}{500} & \multirow{-8}{*}{1.40} \\ \hline
 & train &  & 500 & 2.80 &  &  &  \\ 
 & farsecsq &  & 116 & 12.07 &  &  &  \\ 
 & farsectwo &  & 203 & 6.90 &  &  &  \\ 
 & farsec &  & 470 & 2.98 &  &  &  \\ 
 & clni &  & 440 & 3.18 &  &  &  \\ 
 & clnifarsecsq &  & 71 & 19.72 &  &  &  \\ 
 & clnifarsectwo &  & 151 & 9.27 &  &  &  \\ 
\multirow{-8}{*}{Camel} & clnifarsec & \multirow{-8}{*}{14} & 410 & 3.41 & \multirow{-8}{*}{18} & \multirow{-8}{*}{500} & \multirow{-8}{*}{3.60} \\ \hline
 & train &  & 500 & 9.20 &  &  &  \\ 
 & farsecsq &  & 57 & 80.70 &  &  &  \\ 
 & farsectwo &  & 185 & 24.86 &  &  &  \\ 
 & farsec &  & 489 & 9.41 &  &  &  \\ 
 & clni &  & 446 & 10.31 &  &  &  \\ 
 & clnifarsecsq &  & 48 & 95.83 &  &  &  \\ 
 & clnifarsectwo &  & 168 & 27.38 &  &  &  \\ 
\multirow{-8}{*}{Derby} & clnifarsec & \multirow{-8}{*}{46} & 435 & 10.57 & \multirow{-8}{*}{42} & \multirow{-8}{*}{500} & \multirow{-8}{*}{8.40} \\ \hline
\end{tabular}
\label{tbl:farsecDataset}
\end{table*}

\subsection{Data}

\responseChange{For this work, we compare the differential evolutionary algorithm (DE) and {\IT} to FARSEC using the same data as used in the FARSEC study. The data set includes five projects: four from Apache projects (i.e., Ambari, Camel, Derby and Wicket)~\cite{ohira2015dataset} and one from the Chromium project. For the Apache projects, one thousand bug reports are randomly selected for each project with BUG or IMPROVEMENT label from the JIRA bug tracking system~\cite{ohira2015dataset}. All the selected bug reports are then classified with scripts or manually into six high impact bugs (i.e., \textit{Surprise}, \textit{Dormant}, \textit{Blocking}, \textit{Security}, \textit{Performance}, and \textit{Breakage} bugs). All the target bug reports in our data set all belong to \textit{Security} bug reports (i.e., bug reports of the type \textit{Security}). For the Chromium project, security bugs are labeled as \textit{Bug-Security} when submitted to bug tracking systems. All other types of bug reports in the data set are treated as non-security bug reports.}

\responseChangeTwo{The datasets from FARSEC are publicly available. Our experiments reproduce and improve the FARSEC results using the same datasets.} Table~\ref{tbl:farsecDataset} shows the characteristics of the FARSEC datasets. As we see from the table, one unique feature of the data set is the rarity of the target class. \responseChange{The ``SBRs\%'' column in both training and testing data set indicates that security bug reports make up a very small percentage of the total number of bug reports in projects like Chromium.}

\subsection{Experimental Rig}

\responseChangeTwo{Our experiment design is mainly divided into two parts. When we optimize learners or data pre-processors individually, we divide each {\em training data} into $B=10$ bins, and validate our models using bin $B_i$ after training them on {\em training data - $B_i$}. This 10-fold cross-validation is used to pick the best candidate learner/pre-processor as well as their hyperparameters with the highest performance for that data set. We also need to point out that the 10-fold cross-validation does not apply to the dual optimization, and the way we select the best candidate learner and pre-processor in {\IT} is based on weight calculation and we further refine their hyperparameter's numeric ranges as we discuss in Section~\ref{tion:swift-e}.}

\responseChangeTwo{After finding the best learners and/or pre-processors, we then train the models with the whole training dataset, and test on the separate testing dataset as FARSEC.}

\subsection{Evaluation Metrics}\label{sec:metrics}

To understand the open issues with bug report classification, firstly we must define how they are \textbf{assessed}. If (TN, FN, FP, TP) are the true negatives, false negatives, false positives, and true positives, respectively, found by a classifier, then: 

\bi
\item
{\em pd} = Recall = TP/(TP+FN), the percentage of the actual security bug reports that are predicted to be security bug reports. 
\item
{\em pf} = False Alarms = FP/(FP+TN), the percentage of the non-security bug reports that are reported as security bug reports.
\item
{\em prec} = Precision = TP/(TP+FP), the percentage of the predicted security bug reports that are actual security bug reports. 
\item
\responseChangeTwo{{\em f-score} = F-Measure = 2*pd*prec/(pd+prec), the harmonic mean of the model's precision and recall.} 
\ei

This paper adopts the same evaluation criteria as the original FARSEC paper; i.e. the recall ({\em pd}) and false alarm ({\em pf}) measures. Also, to control the optimization algorithm, we are endeavoring to minimize false alarms while maximizing recall. To achieve those goals, we maximize the {\em g-measure} which is the harmonic mean of recall and the \responseChange{complement} of false alarms in our algorithm.

\begin{equation}\label{eq:one}
 g = \frac{2 \times \mathit{pd} \times (1-\mathit{pf})}{\mathit{pd} + (1-\mathit{pf})}
\end{equation}
{\em g} is maximal when {\em both} recall ({\em pd}) is high and false alarm ({\em pf}) is low. 

\responseChange{We choose {\em g-measure} based on the following considerations. For an imbalanced dataset where there is a skew in the class distribution (e.g., negative samples are much more than positive samples), we have two competing goals:
\bi
\item On the one hand, we want to focus on minimizing false negatives (i.e., security bug reports are not missed in prediction~\cite{scandariato2014predicting}).
\item
On the other hand, we prefer not to predict too many non-security bug reports as security bug reports, which is $(1- pf)$ that also represents specificity.
\ei
}

\responseChange{As to why we use these measures but not some others such as precision, Menzies et al.~\cite{menzies2007problems} argue that when the target class is less than 10\% (as is with all our data), the precision results become more a function of the random number generator used to divide data (for testing purposes). Therefore, we cannot recommend precision for this kind of data. (Aside: we are not alone in this view (that precision should not be used).
For example, the FARSEC paper (that this work builds on) did not assess its models via precision.)}

Besides the above, we also use another evaluation measure called IFA (Initial False Alarm) to evaluate the performance. IFA is the number of initial false alarm encountered before we make the first correct prediction~\cite{huang2017supervised}~\cite{huang2019revisiting}~\cite{huang2019revisiting}. IFA is widely used in defect prediction, and previous works~\cite{kochhar2016practitioners}~\cite{parnin2011automated} have shown that developers are not willing to use a prediction model if the first few recommendations are all false alarms. 

\responseChange{Furthermore, metrics like recall and g-measure are set-based measures, and they are computed using unordered sets of data. To evaluate the results of ranking bug report, mean average precision (MAP) is commonly used to indicate the quality of a ranking by comparing with the ground truth. A higher MAP value usually means more actual security bug reports that predicted are close to the top of the list.}

\responseChange{Equation~\ref{eq:three} and Equation~\ref{eq:four} shows how average precision (AP) and MAP are computed. Specifically, $AP_{n}$ is the average of precision $@k$ where $P(k)$ is the precision at point $k$ in the ranked list and $n$ is the number of predicted security bug reports. As done in the FARSE paper, we say that $MAP_{n}$ is the mean of cumulative average precision scores for each decile.}

\responseChange{\begin{equation}\label{eq:three}
 AP_{n} =\sum_{k=1}^{n}\frac{P(k)}{n}
\end{equation}}

\responseChange{\begin{equation}\label{eq:four}
 MAP_{n} =\sum_{i=1}^{N}\frac{AP_{ni}}{N}
\end{equation}}

\subsection{Statistics} \label{sec:stats}

This study ranks treatments using the Scott-Knott procedure recommended by Mittas \& Angelis in their 2013 IEEE TSE paper~\cite{Mittas13}. This method
sorts results from different treatments, then splits them in order to maximize the expected value of differences  in the observed performances
before and after divisions. For lists $l,m,n$ of size $\mathit{ls},\mathit{ms},\mathit{ns}$ where $l=m\cup n$, the ``best'' division maximizes $E(\Delta)$; i.e. the difference in the expected mean value before and after the spit: 
 \[E(\Delta)=\frac{ms}{ls}abs(m.\mu - l.\mu)^2 + \frac{ns}{ls}abs(n.\mu - l.\mu)^2\]
Scott-Knott then checks if that ``best'' division is actually useful. To implement that check, Scott-Knott would apply some statistical hypothesis test $H$ to check if $m,n$ are significantly different (and if so, Scott-Knott then recurses on each half of the ``best'' division). For this study, our hypothesis test $H$ was a conjunction of the A12 effect size test of and non-parametric bootstrap sampling; i.e. our Scott-Knott divided the data if {\em both} bootstrapping and an effect size test agreed that
the division was statistically significant (95\% confidence) and not a ``small'' effect ($A12 \ge 0.6$).

For a justification of the use of non-parametric bootstrapping, see Efron \& Tibshirani~\cite[p220-223]{efron93}. For a justification of the use of effect size tests see Kampenes~\cite{kampenes2007} who  warn that even if a hypothesis test declares two populations to be ``significantly'' different, then that result is misleading if the ``effect size'' is very small. Hence, to assess the performance differences  we first must rule out small effects. Vargha and Delaney's non-parametric A12 effect size test 
explores two lists $M$ and $N$ of size $m$ and $n$:

\[
A12 = \left(\sum_{x\in M, y \in N} 
\begin{cases} 
1   & \mathit{if}\; x > y\\
0.5 & \mathit{if}\; x == y
\end{cases}\right) / (mn)
\]

This expression computes the probability that the numbers in one sample are bigger than in another. This test was endorsed by Arcuri and Briand~\cite{arcuri2011}. Table~\ref{tbl:resRecall}, Table~\ref{tbl:resFalseAlarm} and Table~\ref{tbl:resIFA} present the results of our Scott-Knott procedure for each project data set. These results are discussed, extensively, in the next section.

\section{Results}
\label{results}

In this section, Table~\ref{tbl:resRecall}, Table~\ref{tbl:resFalseAlarm} and Table~\ref{tbl:resIFA} report results with and without hyperparameter optimization of the pre-processors or learners or both. For the sake of completeness, we also add results of precision and f-measure in Table~\ref{tbl:precisionRes} and Table~\ref{tbl:f-measureRes}. Using those results,
we can now answer our proposed research questions.

\subsection{RQ1}
\begin{RQ}
{\bf RQ1.} \responseChange{Can hyperparameter optimization \responseChangeTwo{techniques} improve the performance of models that better distinguish security bug reports from other bug reports?}
\end{RQ}

\begin{table*} 
\centering
\caption{{\bf RQ1} results: recall. In these results, {\em higher} recalls (a.k.a. pd) are {\em better}. For each row, the best results are \colorbox[HTML]{C0C0C0}{highlighted in gray} (these are the cells that are statistically the same as the best median result -- as judged by our Scott-Knot test). Across all rows, {\IT} has the most number of best results.}
\small
\begin{tabular}{l|l|c|c|c|c|c}
\multicolumn{2}{c}{} & \begin{tabular}[c]{@{}c@{}}Prior state\\ of the art\\ ~\cite{peters2018text}\end{tabular} & \begin{tabular}[c]{@{}c@{}}Optimize\\ learners\\ (only)\end{tabular} & \begin{tabular}[c]{@{}c@{}}Data\\ pre-processing\\ (no tuning)\end{tabular} & \begin{tabular}[c]{@{}c@{}}Data\\ pre-processing\\ (tuned)\end{tabular} & \begin{tabular}[c]{@{}c@{}}Tune both\\ (dual)\end{tabular} \\ \hline
\rowcolor[HTML]{ECF4FF} 
\multicolumn{1}{c|}{\cellcolor[HTML]{ECF4FF}\textbf{Project}} & \multicolumn{1}{c|}{\cellcolor[HTML]{ECF4FF}\textbf{Filter}} & \textbf{FARSEC} & \textbf{\begin{tabular}[c]{@{}c@{}}DE+\\ Learners\end{tabular}} & \textbf{Pre-processors} & \textbf{\begin{tabular}[c]{@{}c@{}}DE+\\ Pre-processors\end{tabular}} & \textbf{SWIFT} \\ \hline

 & train & 15.7 & 46.9 & 68.7 & 73.9 & \cellcolor[HTML]{C0C0C0}86.1 \\  
 & farsecsq & 14.8 & 64.3 & 80.0 & \cellcolor[HTML]{C0C0C0}84.3 & 72.2 \\  
 & farsectwo & 15.7 & 40.9 & \cellcolor[HTML]{C0C0C0}78.3 & 77.4 & 77.4 \\  
 & farsec & 15.7 & 46.1 & \cellcolor[HTML]{C0C0C0}80.8 & 72.2 & 77.4 \\  
 & clni & 15.7 & 30.4 & 74.8 & 72.2 & \cellcolor[HTML]{C0C0C0}80.9 \\  
 & clnifarsecsq & 49.6 & 72.2 & 82.6 & \cellcolor[HTML]{C0C0C0}86.1 & 72.2 \\ 
 & clnifarsectwo & 15.7 & 50.4 & \cellcolor[HTML]{C0C0C0}79.1 & 74.8 & 78.3 \\
\multirow{-8}{*}{Chromium} & clnifarsec & 15.7 & 47.8 & \cellcolor[HTML]{C0C0C0}78.3 & 74.7 & 72.2 \\\cline{2-7} 
\multicolumn{2}{r|}{\textit{Median Recall}} & 15.7 & 47.3 & 78.7 & 74.8 & 77.4 \\
 \multicolumn{7}{r}{~}\\\hline
 & train & 16.7 & 0.0 & \cellcolor[HTML]{C0C0C0}66.7 & \cellcolor[HTML]{C0C0C0}66.7 & 50.0 \\  
 & farsecsq & 66.7 & 50.0 & \cellcolor[HTML]{C0C0C0}83.3 & \cellcolor[HTML]{C0C0C0}83.3 & \cellcolor[HTML]{C0C0C0}83.3 \\  
 & farsectwo & \cellcolor[HTML]{C0C0C0}66.7 & 50.0 & \cellcolor[HTML]{C0C0C0}66.7 & \cellcolor[HTML]{C0C0C0}66.7 & \cellcolor[HTML]{C0C0C0}66.7 \\  
 & farsec & 33.3 & \cellcolor[HTML]{C0C0C0}66.7 & \cellcolor[HTML]{C0C0C0}66.7 & \cellcolor[HTML]{C0C0C0}66.7 & \cellcolor[HTML]{C0C0C0}66.7 \\  
 & clni & 0.0 & 16.7 & \cellcolor[HTML]{C0C0C0}50.0 & \cellcolor[HTML]{C0C0C0}50.0 & \cellcolor[HTML]{C0C0C0}50.0 \\  
 & clnifarsecsq & 33.3 & \cellcolor[HTML]{C0C0C0}83.3 & \cellcolor[HTML]{C0C0C0}83.3 & \cellcolor[HTML]{C0C0C0}83.3 & \cellcolor[HTML]{C0C0C0}83.3 \\  
 & clnifarsectwo & 33.3 & 50.0 & \cellcolor[HTML]{C0C0C0}66.7 & \cellcolor[HTML]{C0C0C0}66.7 & \cellcolor[HTML]{C0C0C0}66.7 \\  
\multirow{-8}{*}{Wicket} & clnifarsec & 50.0 & \cellcolor[HTML]{C0C0C0}66.7 & \cellcolor[HTML]{C0C0C0}66.7 & \cellcolor[HTML]{C0C0C0}66.7 & \cellcolor[HTML]{C0C0C0}66.7 \\\cline{2-7} 
\multicolumn{2}{r|}{\textit{Median Recall}} & 33.3 & 50.0 & 66.7 & 66.7 & 66.7\\
 \multicolumn{7}{r}{~}\\\hline
 & train & 14.3 & 28.6 & 57.1 & 57.1 & \cellcolor[HTML]{C0C0C0}85.7 \\  
 & farsecsq & 42.9 & 57.1 & 57.1 & 57.1 & \cellcolor[HTML]{C0C0C0}85.7 \\  
 & farsectwo & 57.1 & 57.1 & 57.1 & 57.1 & \cellcolor[HTML]{C0C0C0}85.7 \\  
 & farsec & 14.3 & 57.1 & 57.1 & 57.1 & \cellcolor[HTML]{C0C0C0}85.7 \\  
 & clni & 14.3 & 28.6 & 57.1 & 57.1 & \cellcolor[HTML]{C0C0C0}85.7 \\  
 & clnifarsecsq & 57.1 & 57.1 & 57.1 & 57.1 & \cellcolor[HTML]{C0C0C0}71.4 \\ 
 & clnifarsectwo & 28.6 & 57.1 & 57.1 & 57.1 & \cellcolor[HTML]{C0C0C0}85.7 \\
\multirow{-8}{*}{Ambari} & clnifarsec & 14.3 & 57.1 & 57.1 & 57.1 & \cellcolor[HTML]{C0C0C0}85.7 \\\cline{2-7}
\multicolumn{2}{r|}{\textit{Median Recall}} & 21.5 & 57.1 & 57.1 & 57.1 & 85.7\\
 \multicolumn{7}{r}{~}\\\hline
 & train & 11.1 & 16.7 & 33.3 & 44.4 & \cellcolor[HTML]{C0C0C0}55.6 \\  
 & farsecsq & 16.7 & 44.4 & 44.4 & 55.6 & \cellcolor[HTML]{C0C0C0}66.7 \\  
 & farsectwo & 50.0 & 44.4 & \cellcolor[HTML]{C0C0C0}61.1 & \cellcolor[HTML]{C0C0C0}61.1 & \cellcolor[HTML]{C0C0C0}61.1 \\  
 & farsec & 16.7 & 22.2 & 33.3 & 33.3 & \cellcolor[HTML]{C0C0C0}55.6 \\  
 & clni & 16.7 & 16.7 & 33.3 & 38.9 & \cellcolor[HTML]{C0C0C0}50.0 \\  
 & clnifarsecsq & 16.7 & 38.9 & 27.8 & 33.3 & \cellcolor[HTML]{C0C0C0}61.1 \\ 
 & clnifarsectwo & 11.1 & 61.1 & \cellcolor[HTML]{C0C0C0}72.2 & 61.1 & 61.1 \\
\multirow{-8}{*}{Camel} & clnifarsec & 16.7 & 22.2 & 33.3 & 38.9 & \cellcolor[HTML]{C0C0C0}55.6 \\\cline{2-7}
\multicolumn{2}{r|}{\textit{Median Recall}} & 16.7 & 38.5 & 33.3 & 41.7 & 58.4\\
 \multicolumn{7}{r}{~}\\\hline
 & train & 38.1 & 47.6 & 54.7 & 59.5 & \cellcolor[HTML]{C0C0C0}69.0 \\  
 & farsecsq & 54.8 & 59.5 & 54.7 & \cellcolor[HTML]{C0C0C0}66.7 & \cellcolor[HTML]{C0C0C0}66.7 \\  
 & farsectwo & 47.6 & 59.5 & 47.6 & 66.7 & \cellcolor[HTML]{C0C0C0}78.6 \\  
 & farsec & 38.1 & 47.6 & 57.1 & 59.5 & \cellcolor[HTML]{C0C0C0}64.3 \\  
 & clni & 23.8 & 45.2 & 57.7 & 61.9 & \cellcolor[HTML]{C0C0C0}69.0 \\  
 & clnifarsecsq & 54.8 & 59.5 & 76.2 & \cellcolor[HTML]{C0C0C0}69.0 & 66.7 \\ 
 & clnifarsectwo & 35.7 & 59.5 & 54.8 & 61.9 & \cellcolor[HTML]{C0C0C0}66.7 \\
\multirow{-8}{*}{Derby} & clnifarsec & 38.1 & 47.6 & 61.9 & 57.1 & \cellcolor[HTML]{C0C0C0}66.7 \\\cline{2-7} 
\multicolumn{2}{r|}{\textit{Median Recall}} & 38.1 & 53.6 & 56.0 & 61.9 & 66.7\\
\multicolumn{7}{r}{~}\\\hline 
\rowcolor[HTML]{ECF4FF}\multicolumn{2}{r|}{\textit{Overall Median Recall}} & 21.5 & 50.0 & 57.1 & 61.9 & 66.7\\\hline
\end{tabular}
\label{tbl:resRecall}
\end{table*}
\subsubsection{Recall Results}

\responseChange{In the recall results of Table~\ref{tbl:resRecall}, we can observe that FARSEC rarely achieves the best results while {\IT} is much better than FARSEC. For example:
\bi
\item In the Chromium project, median recall changes from 15.7\% to 77.4\% from FARSEC to {\IT}.
\item In the Ambari project, the median recall changes from 21.5\% to 85.7\% from FARSEC to {\IT}. 
\item Overall, as shown in the last line of Table~\ref{tbl:resRecall}, the improvement is from 21.5\% to 66.7\% (FARSEC to {\IT}).
\ei
}

\responseChange{In addition, in Table~\ref{tbl:resRecall},  the gray cells show  the ``best'' results in each row (where ``best'' is defined using the statistical significance tests of  Section~\ref{sec:stats}). Overall, {\IT} is statistically significantly best in 31/40 of all the rows of Table~\ref{tbl:resRecall}. This is more than twice as many wins as other approaches explored in this table; e.g. DE+pre-processors scores best in only 13/40 rows. Hence, for this data set, we say that dual optimization of both learners and pre-processors work best.} 
 
\responseChange{Just for completeness, we note that for all methods with any data pre-processing procedure (i.e., in the last three columns of Table~\ref{tbl:resRecall}) work well for the Wicket project. Clearly, for this data set, data pre-processing such as repairing the class imbalance issue is essential for good performance.}

\begin{table*}
\centering
\caption{{\bf RQ1} results: false positive rate (a.k.a., pf), the lower values are better. Same as Table~\ref{tbl:resRecall}; i.e. the best results are highlighted in grey cells. While FARSEC has the most best results, these low false positive rates are only achieved by settling for low recalls (see Table~\ref{tbl:resRecall}).}
\small
\begin{tabular}{l|l|c|c|c|c|c}

\multicolumn{2}{c}{} & \begin{tabular}[c]{@{}c@{}}Prior state\\ of the art\\ ~\cite{peters2018text}\end{tabular} & \begin{tabular}[c]{@{}c@{}}Optimize\\ learners\\ (only)\end{tabular} & \begin{tabular}[c]{@{}c@{}}Data\\ pre-processing\\ (no tuning)\end{tabular} & \begin{tabular}[c]{@{}c@{}}Data\\ pre-processing\\ (tuned)\end{tabular} & \begin{tabular}[c]{@{}c@{}}Tune both\\ (dual)\end{tabular} \\ \hline
\rowcolor[HTML]{ECF4FF} 
\multicolumn{1}{c|}{\cellcolor[HTML]{ECF4FF}\textbf{Project}} & \multicolumn{1}{c|}{\cellcolor[HTML]{ECF4FF}\textbf{Filter}} & \textbf{FARSEC} & \textbf{\begin{tabular}[c]{@{}c@{}}DE+\\ Learners\end{tabular}} & \textbf{Pre-processors} & \textbf{\begin{tabular}[c]{@{}c@{}}DE+\\ Pre-processors\end{tabular}} & \textbf{SWIFT} \\ \hline
 & train & \cellcolor[HTML]{C0C0C0}0.2 & 6.8 & 24.1 & 17.8 & 24.0 \\  
 & farsecsq & \cellcolor[HTML]{C0C0C0}0.3 & 10.3 & 31.5 & 25.1 & 14.3 \\  
 & farsectwo & \cellcolor[HTML]{C0C0C0}0.2 & 6.5 & 27.6 & 23.1 & 26.1 \\  
 & farsec & \cellcolor[HTML]{C0C0C0}0.2 & 6.9 & 36.1 & 14.9 & 14.7 \\  
 & clni & \cellcolor[HTML]{C0C0C0}0.2 & 4.1 & 24.8 & 13.6 & 26.2 \\  
 & clnifarsecsq & \cellcolor[HTML]{C0C0C0}3.8 & 14.2 & 30.4 & 25.6 & 14.0 \\  
 & clnifarsectwo & \cellcolor[HTML]{C0C0C0}0.2 & 7.0 & 29.9 & 12.8 & 18.9 \\  
\multirow{-8}{*}{Chromium} & clnifarsec & \cellcolor[HTML]{C0C0C0}0.2 & 10.4 & 29.0 & 17.1 & 20.2 \\ \cline{2-7}
\multicolumn{2}{r|}{\textit{Median FPR}} & 0.2 & 7.0 & 29.5 & 17.5 & 19.5\\
\multicolumn{7}{r}{~}\\\hline
 & train & 7.1 & \cellcolor[HTML]{C0C0C0}5.1 & 32.0 & 12.1 & 27.5 \\  
 & farsecsq & \cellcolor[HTML]{C0C0C0}38.3 & 44.5 & 71.3 & 66.8 & 66.7 \\  
 & farsectwo & \cellcolor[HTML]{C0C0C0}36.6 & 42.3 & 68.2 & 62.9 & 61.5 \\  
 & farsec & \cellcolor[HTML]{C0C0C0}8.1 & 23.1 & 43.9 & 26.1 & 23.3 \\  
 & clni & 5.5 & \cellcolor[HTML]{C0C0C0}2.4 & 21.1 & 12.5 & 14.4 \\  
 & clnifarsecsq & \cellcolor[HTML]{C0C0C0}25.5 & 66.8 & 66.8 & 66.8 & 57.5 \\
 & clnifarsectwo & \cellcolor[HTML]{C0C0C0}27.7 & 39.9 & 61.3 & 61.3 & 52.8 \\
\multirow{-8}{*}{Wicket} & clnifarsec & \cellcolor[HTML]{C0C0C0}10.5 & 23.1 & 38.9 & 22.9 & 22.1 \\ \cline{2-7}
\multicolumn{2}{r|}{\textit{Median FPR}} & 18.0 & 31.5 & 52.6 & 43.7 & 40.2\\
\multicolumn{7}{r}{~}\\\hline
 & train & 1.6 & \cellcolor[HTML]{C0C0C0}0.8 & 20.1 & 10.8 & 17.8 \\  
 & farsecsq & 14.4 & \cellcolor[HTML]{C0C0C0}2.8 & 30.4 & 17.2 & 23.7 \\  
 & farsectwo & 3.0 & \cellcolor[HTML]{C0C0C0}2.8 & 22.1 & 17.8 & 19.7 \\  
 & farsec & 4.9 & \cellcolor[HTML]{C0C0C0}2.0 & 19.9 & 7.1 & 20.3 \\  
 & clni & 2.6 & \cellcolor[HTML]{C0C0C0}0.8 & 12.4 & 8.9 & 18.1 \\  
 & clnifarsecsq & 7.7 & \cellcolor[HTML]{C0C0C0}2.4 & 13.4 & 7.1 & 29.0 \\  
 & clnifarsectwo & 4.5 & \cellcolor[HTML]{C0C0C0}2.8 & 13.0 & 5.1 & 22.7 \\  
\multirow{-8}{*}{Ambari} & clnifarsec & \cellcolor[HTML]{C0C0C0}0.0 & 2.4 & 7.9 & 3.9 & 18.9 \\ \cline{2-7}
\multicolumn{2}{r|}{\textit{Median FPR}} & 3.8 & 2.4 & 16.7 & 8.0 & 20.0\\
\multicolumn{7}{r}{~}\\\hline
 & train & 3.5 & \cellcolor[HTML]{C0C0C0}1.5 & 27.4 & 35.9 & 15.8 \\  
 & farsecsq & \cellcolor[HTML]{C0C0C0}11.4 & 24.7 & 20.5 & 23.4 & 27.8 \\  
 & farsectwo & 41.8 & \cellcolor[HTML]{C0C0C0}17.6 & 71.0 & 53.1 & 45.2 \\  
 & farsec & \cellcolor[HTML]{C0C0C0}6.9 & 12.4 & 39.4 & 28.0 & 35.7 \\  
 & clni & 12.3 & \cellcolor[HTML]{C0C0C0}7.9 & 33.6 & 35.3 & 24.7 \\  
 & clnifarsecsq & 13.9 & 14.9 & \cellcolor[HTML]{C0C0C0}12.4 & 15.6 & 27.2 \\
 & clnifarsectwo & \cellcolor[HTML]{C0C0C0}7.7 & 50.0 & 64.9 & 51.9 & 38.8 \\
\multirow{-8}{*}{Camel} & clnifarsec & \cellcolor[HTML]{C0C0C0}5.0 & 11.6 & 24.9 & 34.4 & 37.1 \\ \cline{2-7}
\multicolumn{2}{r|}{\textit{Median FPR}} & 9.6 & 13.7 & 30.5 & 34.8 & 31.8\\
\multicolumn{7}{r}{~}\\\hline
 & train & \cellcolor[HTML]{C0C0C0}6.8 & 39.3 & 22.2 & 20.7 & 19.7 \\  
 & farsecsq & 29.9 & 40.6 & 51.7 & 51.5 & \cellcolor[HTML]{C0C0C0}22.5 \\  
 & farsectwo & \cellcolor[HTML]{C0C0C0}12.4 & 24.2 & 27.9 & 33.6 & 40.0 \\  
 & farsec & 6.3 & \cellcolor[HTML]{C0C0C0}4.1 & 21.0 & 19.0 & 13.8 \\  
 & clni & \cellcolor[HTML]{C0C0C0}0.4 & 3.5 & 16.8 & 24.5 & 25.5 \\  
 & clnifarsecsq & \cellcolor[HTML]{C0C0C0}29.9 & 42.4 & 74.7 & 65.1 & 42.3 \\ 
 & clnifarsectwo & \cellcolor[HTML]{C0C0C0}9.2 & 24.2 & 36.5 & 30.3 & 52.2 \\ 
\multirow{-8}{*}{Derby} & clnifarsec & 6.8 & \cellcolor[HTML]{C0C0C0}3.9 & 28.8 & 10.9 & 19.6 \\ \cline{2-7}
\multicolumn{2}{r|}{\textit{Median FPR}} & 8.0 & 24.2 & 28.4 & 27.4 & 24.0\\
\multicolumn{7}{r}{~}\\\hline 
\rowcolor[HTML]{ECF4FF}\multicolumn{2}{r|}{\textit{Overall Median FPR}} & 8.0 & 13.7 & 29.5 & 27.4 & 24.0\\\hline
\end{tabular}
\label{tbl:resFalseAlarm}
\end{table*}

\subsubsection{False Positive Rate Results}\label{tion:fp}

\responseChange{As to the false positive rate results, Table~\ref{tbl:resFalseAlarm} shows that FARSEC has the lowest false positive rate across more than half of the datasets with filters. However, as shown in Table~\ref{tbl:resRecall}, FARSEC achieves those low false positive rate by settling for some low recalls.}

\responseChange{As to {\IT}, we note that its improvements in recall (seen above) come at the cost of some increments in false positive rate. As shown in the last line of Table~\ref{tbl:resFalseAlarm}, the overall median false positive rate increases from 8\% to 24\% (FARSEC to SWIFT). While, ideally, the false positive rate is zero, it is inevitable that there is some cost in dealing with security problems. Another way to look at this is to say while our methods help distinguishing security bug reports (from other bug reports), they also highlight the costs involved in securing software. Our method can better distinguish security bug reports than the prior state-of-the-art. However, to do so, there is some increase in the workload of developers who have to read more code and suffer a (slightly) higher false positive rate. Such is the price of software quality assurance.}

\responseChange{Hence we say that this 16\% increase in overall false positive rates is the {\em acceptable} and {\em inevitable} ``price'' of increasing recall. As to {\em acceptable}, the overall false alarms are still less than a quarter -- which is in the same range as many other software analytic applications\footnote{e.g. Figure~12 of~\cite{menzies2007data} lists nine SE data mining applications with median false positive rates of 25\%).}. }

\responseChange{As to {\em inevitable}, consider two models:
\bi
\item One just predicts ``yes'' all the time. This model has 100\% recall (since it finds every target class) but it suffers from large false positive rates.
\item
Another model just predicts ``no'' all the time. This second model has 0\% false positive rate (i.e., it never makes mistakes in prediction) but it also has a 0\% recall (since it never finds any target class).
\ei
}

\responseChange{In practice, all learners make trade-offs between recall and false positive rate as they explore models somewhere on a curve between:
\bi
\item
\responseChangeTwo{Recall from 0\% to 100\%}
\item
\responseChangeTwo{False positive rate from 0\% to 100\%}
\item
In addition, unless the learner is broken, this curve bends upwards away from the \mbox{recall == false positive rate} line towards the point recall=100\% and false positive rate=0\% (but rarely does any learner reach this point). 
\ei
This means that as a learner tries different models, increased recall comes at the cost of also increasing false positive rates. The trick here is to increase recall {\em more than} false positive rate, as is done by {\IT}. In this paper, we show that we can increase median recall from 21.5\% to 66.7\% (while at the same time only increasing median false positive rate by 16\% from 8\% to 24\%).
~\\
}

\begin{table*}
\centering
\caption{{\bf RQ1} results: initial false alarm (IFA). IFA is the number of false alarms developers must suffer through before finding their first target. Lower values are better. Same as format as Table~\ref{tbl:resRecall}; i.e. best results are shown in grey.}
\begin{tabular}{l|l|c|c|c|c|c}

\multicolumn{2}{c}{} & \begin{tabular}[c]{@{}c@{}}Prior state\\ of the art\\ ~\cite{peters2018text}\end{tabular} & \begin{tabular}[c]{@{}c@{}}Optimize\\ learners\\ (only)\end{tabular} & \begin{tabular}[c]{@{}c@{}}Data\\ pre-processing\\ (no tuning)\end{tabular} & \begin{tabular}[c]{@{}c@{}}Data\\ pre-processing\\ (tuned)\end{tabular} & \begin{tabular}[c]{@{}c@{}}Tune both\\ (dual)\end{tabular} \\ \hline
\rowcolor[HTML]{ECF4FF} 
\multicolumn{1}{c|}{\cellcolor[HTML]{ECF4FF}\textbf{Project}} & \multicolumn{1}{c|}{\cellcolor[HTML]{ECF4FF}\textbf{Filter}} & \textbf{FARSEC} & \textbf{\begin{tabular}[c]{@{}c@{}}DE+\\ Learners\end{tabular}} & \textbf{Pre-processors} & \textbf{\begin{tabular}[c]{@{}c@{}}DE+\\ Pre-processors\end{tabular}} & \textbf{SWIFT} \\ \hline
 & train & N/A & 62 & 75 & 61 & \cellcolor[HTML]{C0C0C0}58 \\  
 & farsecsq & N/A & \cellcolor[HTML]{C0C0C0}20 & 72 & 54 & 36 \\  
 & farsectwo & N/A & \cellcolor[HTML]{C0C0C0}37 & 91 & 78 & 87 \\  
 & farsec & N/A & 62 & 112 & 62 & \cellcolor[HTML]{C0C0C0}56 \\  
 & clni & N/A & \cellcolor[HTML]{C0C0C0}41 & 86 & 48 & 74 \\  
 & clnifarsecsq & N/A & 41 & 57 & 62 & \cellcolor[HTML]{C0C0C0}37 \\  
 & clnifarsectwo & N/A & \cellcolor[HTML]{C0C0C0}37 & 89 & 47 & 58 \\
\multirow{-8}{*}{Chromium} & clnifarsec & N/A & 62 & 113 & 63 & \cellcolor[HTML]{C0C0C0}54 \\ \cline{2-7}
\multicolumn{2}{r|}{\textit{Median IFA}} & N/A & 41 & 88 & 62 & 57\\
\multicolumn{7}{r}{~}\\\hline
 & train & N/A & \cellcolor[HTML]{C0C0C0}25 & 60 & 34 & 46 \\  
 & farsecsq & N/A & \cellcolor[HTML]{C0C0C0} 29 & 37 & 33 & 39 \\  
 & farsectwo & N/A & 32 & 35 & 34 & \cellcolor[HTML]{C0C0C0}31 \\  
 & farsec & N/A & 23 & 44 & 30 & \cellcolor[HTML]{C0C0C0}22 \\  
 & clni & N/A & \cellcolor[HTML]{C0C0C0}12 & 44 & 21 & 27 \\  
 & clnifarsecsq & N/A & 9 & 8 & 9 & \cellcolor[HTML]{C0C0C0}6 \\  
 & clnifarsectwo & N/A & \cellcolor[HTML]{C0C0C0}8 & 11 & 12 & \cellcolor[HTML]{C0C0C0}8 \\  
\multirow{-8}{*}{Wicket} & clnifarsec & N/A & 17 & 33 & \cellcolor[HTML]{C0C0C0}15 & 18 \\ \cline{2-7}
\multicolumn{2}{r|}{\textit{Median IFA}} & N/A & 20 & 36 & 26 & 25\\
\multicolumn{7}{r}{~}\\\hline
 & train & N/A & 7 & 8 & 9 & \cellcolor[HTML]{C0C0C0}4 \\  
 & farsecsq & N/A & 8 & 21 & 14 & \cellcolor[HTML]{C0C0C0}7 \\  
 & farsectwo & N/A & \cellcolor[HTML]{C0C0C0}1 & 19 & 12 & 3 \\  
 & farsec & N/A & \cellcolor[HTML]{C0C0C0}1 & 35 & 24 & 17 \\  
 & clni & N/A & \cellcolor[HTML]{C0C0C0}1 & 32 & 19 & 13 \\  
 & clnifarsecsq & N/A & \cellcolor[HTML]{C0C0C0}8 & 18 & 10 & \cellcolor[HTML]{C0C0C0}8 \\  
 & clnifarsectwo & N/A & \cellcolor[HTML]{C0C0C0}7 & 28 & 8 & 11 \\  
\multirow{-8}{*}{Ambari} & clnifarsec & N/A & 5 & 10 & \cellcolor[HTML]{C0C0C0}4 & 17 \\ \cline{2-7}
\multicolumn{2}{r|}{\textit{Median IFA}} & N/A & 6 & 20 & 11 & 10\\
\multicolumn{7}{r}{~}\\\hline
 & train & N/A & \cellcolor[HTML]{C0C0C0}6 & 19 & 23 & 15 \\  
 & farsecsq & N/A & 23 & 29 & 32 & \cellcolor[HTML]{C0C0C0}14 \\  
 & farsectwo & N/A & \cellcolor[HTML]{C0C0C0}4 & 13 & 8 & 25 \\  
 & farsec & N/A & 17 & 21 & 20 & \cellcolor[HTML]{C0C0C0}8 \\  
 & clni & N/A & \cellcolor[HTML]{C0C0C0}16 & 37 & 33 & 30 \\  
 & clnifarsecsq & N/A & 5 & \cellcolor[HTML]{C0C0C0}3 & \cellcolor[HTML]{C0C0C0}3 & 4 \\  
 & clnifarsectwo & N/A & 19 & 22 & 15 & \cellcolor[HTML]{C0C0C0}12 \\ 
\multirow{-8}{*}{Camel} & clnifarsec & N/A & \cellcolor[HTML]{C0C0C0}14 & 23 & 29 & 22 \\ \cline{2-7}
\multicolumn{2}{r|}{\textit{Median IFA}} & N/A & 15 & 22 & 22 & 15\\
\multicolumn{7}{r}{~}\\\hline
 & train & N/A & 4 & 6 & 3 & \cellcolor[HTML]{C0C0C0}2 \\  
 & farsecsq & N/A & \cellcolor[HTML]{C0C0C0}4 & \cellcolor[HTML]{C0C0C0}4 & \cellcolor[HTML]{C0C0C0}4 & \cellcolor[HTML]{C0C0C0}4 \\  
 & farsectwo & N/A & 4 & \cellcolor[HTML]{C0C0C0}3 & 5 & \cellcolor[HTML]{C0C0C0}3 \\  
 & farsec & N/A & \cellcolor[HTML]{C0C0C0}1 & 8 & 7 & 4 \\  
 & clni & N/A & \cellcolor[HTML]{C0C0C0}1 & 8 & 5 & 3 \\  
 & clnifarsecsq & N/A & \cellcolor[HTML]{C0C0C0}1 & 2 & 2 & \cellcolor[HTML]{C0C0C0}1 \\  
 & clnifarsectwo & N/A & \cellcolor[HTML]{C0C0C0}2 & 9 & 8 & 4 \\    
\multirow{-8}{*}{Derby} & clnifarsec & N/A & \cellcolor[HTML]{C0C0C0}1 & 3 & 3 & 2 \\ \cline{2-7}
\multicolumn{2}{r|}{\textit{Median IFA}} & N/A & 2 & 5 & 5 & 3\\
\multicolumn{7}{r}{~}\\\hline 
\rowcolor[HTML]{ECF4FF}\multicolumn{2}{r|}{\textit{Overall median IFA}} & N/A & 15 & 22 & 22 & 15\\\hline
\end{tabular}
\label{tbl:resIFA}
\end{table*}

\subsubsection{Initial False Alarms Results}

IFA is the number of false positives a programmer must suffer through before they find a real security bug report. Table~\ref{tbl:resIFA} shows our IFA results. There are three points to note from this table:
\bi
\item FARSEC has no results in this table because FARSEC does not report results for this metric.
\item For IFA, methods that only with/tune the data pre-processors perform worse than methods that optimize the learners (i.e., DE+Learners and {\IT}).
\item In terms of absolute numbers, the IFA results are low for the Derby project. From \tbl{farsecDataset}, we can conjecture a reason for this -- of the data set with a higher percentage of security bug reports, the data sets have the more known target class, which is more likely to reduce the number of false positives encounter before the first correct prediction. 
\item At the other end of the spectrum, IFA is much larger for the Chromium project (median values for DE+learner or {\IT} of about 40 or 60). This result highlights the high cost of building highly secure software. When the target class is rare, even with our best-of-breed methods, some non-trivial amount of manual effort may be required.
\ei

\begin{table*}[!htbp]
\centering
\caption{{\bf RQ1} results: precision. Higher values are better. Same as format as Table~\ref{tbl:resRecall}; i.e. best results are shown in grey.}
\begin{tabular}{l|l|c|c|c|c|c}
\multicolumn{2}{c}{} & \begin{tabular}[c]{@{}c@{}}Prior state\\ of the art\\ ~\cite{peters2018text}\end{tabular} & \begin{tabular}[c]{@{}c@{}}Optimize\\ learners\\ (only)\end{tabular} & \begin{tabular}[c]{@{}c@{}}Data\\ pre-processing\\ (no tuning)\end{tabular} & \begin{tabular}[c]{@{}c@{}}Data\\ pre-processing\\ (tuned)\end{tabular} & \begin{tabular}[c]{@{}c@{}}Tune both\\ (dual)\end{tabular} \\ \hline
\rowcolor[HTML]{ECF4FF} 
\multicolumn{1}{c|}{\cellcolor[HTML]{ECF4FF}\textbf{Project}} & \multicolumn{1}{c|}{\cellcolor[HTML]{ECF4FF}\textbf{Filter}} & \textbf{FARSEC} & \textbf{\begin{tabular}[c]{@{}c@{}}DE+\\ Learners\end{tabular}} & \textbf{Pre-processors} & \textbf{\begin{tabular}[c]{@{}c@{}}DE+\\ Pre-processors\end{tabular}} & \textbf{SWIFT} \\ \hline
 & train & \cellcolor[HTML]{C0C0C0}31.0 & 3.6 & 1.5 & 2.2 & 1.9 \\  
 & farsecsq & \cellcolor[HTML]{C0C0C0}23.9 & 3.3 & 1.4 & 1.8 & 2.7 \\  
 & farsectwo & \cellcolor[HTML]{C0C0C0}31.0 & 3.4 & 1.5 & 1.8 & 1.6 \\  
 & farsec & \cellcolor[HTML]{C0C0C0}31.0 & 3.6 & 1.2 & 2.6 & 2.8 \\  
 & clni & \cellcolor[HTML]{C0C0C0}27.7 & 3.8 & 1.6 & 2.8 & 1.7 \\  
 & clnifarsecsq & \cellcolor[HTML]{C0C0C0}6.7 & 2.7 & 1.5 & 1.8 & 2.8 \\  
 & clnifarsectwo & \cellcolor[HTML]{C0C0C0}27.7 & 3.8 & 1.4 & 3.1 & 2.2 \\  
\multirow{-8}{*}{Chromium} & clnifarsec & \cellcolor[HTML]{C0C0C0}27.7 & 2.4 & 1.5 & 2.3 & 1.9 \\ \cline{2-7}
\multicolumn{2}{r|}{\textit{Median Prec}} & 27.7 & 3.5 & 1.5 & 2.3 & 2.1\\
\multicolumn{7}{r}{~}\\\hline
 & train & 2.8 & 0.0 & 2.5 & \cellcolor[HTML]{C0C0C0}6.3 & 2.2 \\  
 & farsecsq & \cellcolor[HTML]{C0C0C0}2.1 & 1.4 & 1.4 & 1.5 & 1.5 \\  
 & farsectwo & \cellcolor[HTML]{C0C0C0}2.2 & 1.4 & 1.2 & 1.3 & 1.3 \\  
 & farsec & \cellcolor[HTML]{C0C0C0}4.8 & 3.4 & 1.8 & 3.0 & 3.4 \\  
 & clni & 0.0 & \cellcolor[HTML]{C0C0C0}8.3 & 2.8 & 4.7 & 4.1 \\  
 & clnifarsecsq & \cellcolor[HTML]{C0C0C0}1.6 & 1.2 & 1.2 & 1.2 & 1.4 \\  
 & clnifarsectwo & 1.4 & \cellcolor[HTML]{C0C0C0}1.5 & 1.3 & 1.3 & \cellcolor[HTML]{C0C0C0}1.5 \\  
\multirow{-8}{*}{Wicket} & clnifarsec & \cellcolor[HTML]{C0C0C0}5.5 & 3.4 & 2.0 & 3.4 & 3.5 \\ \cline{2-7}
\multicolumn{2}{r|}{\textit{Median Prec}} & 2.2 & 1.5 & 1.6 & 2.3 & 1.9\\
\multicolumn{7}{r}{~}\\\hline
 & train & 11.1 & \cellcolor[HTML]{C0C0C0}40.0 & 2.9 & 5.4 & 5.4 \\  
 & farsecsq & 4.1 & \cellcolor[HTML]{C0C0C0}18.8 & 2.0 & 3.4 & 4.1 \\  
 & farsectwo & \cellcolor[HTML]{C0C0C0}21.1 & 18.8 & 2.7 & 3.3 & 4.9 \\  
 & farsec & 4.0 & \cellcolor[HTML]{C0C0C0}25.0 & 3.0 & 7.9 & 4.8 \\  
 & clni & 7.1 & \cellcolor[HTML]{C0C0C0}40.0 & 4.7 & 6.5 & 5.3 \\  
 & clnifarsecsq & 9.5 & \cellcolor[HTML]{C0C0C0}21.4 & 4.3 & 7.9 & 2.7 \\  
 & clnifarsectwo & 8.3 & \cellcolor[HTML]{C0C0C0}18.8 & 4.5 & 10.7 & 4.3 \\  
\multirow{-8}{*}{Ambari} & clnifarsec & \cellcolor[HTML]{C0C0C0}100.0 & 21.4 & 7.3 & 13.6 & 5.1 \\ \cline{2-7}
\multicolumn{2}{r|}{\textit{Median Prec}} & 8.9 & 21.4 & 3.7 & 7.2 & 4.9\\
\multicolumn{7}{r}{~}\\\hline
 & train & 10.5 & \cellcolor[HTML]{C0C0C0}30.0 & 3.6 & 3.9 & 11.6 \\  
 & farsecsq & 5.2 & 5.6 & 6.7 & 8.2 & \cellcolor[HTML]{C0C0C0}8.3 \\  
 & farsectwo & 4.3 & \cellcolor[HTML]{C0C0C0}7.7 & 2.8 & 3.8 & 4.4 \\  
 & farsec & \cellcolor[HTML]{C0C0C0}8.3 & 4.8 & 2.6 & 3.6 & 5.5 \\  
 & clni & 4.8 & \cellcolor[HTML]{C0C0C0}7.3 & 3.0 & 4.0 & 7.0 \\  
 & clnifarsecsq & 4.3 & \cellcolor[HTML]{C0C0C0}9.0 & 7.8 & 6.2 & 7.1 \\  
 & clnifarsectwo & \cellcolor[HTML]{C0C0C0}5.1 & 4.0 & 3.7 & 3.8 & \cellcolor[HTML]{C0C0C0}5.1 \\  
\multirow{-8}{*}{Camel} & clnifarsec & \cellcolor[HTML]{C0C0C0}11.1 & 5.2 & 4.0 & 4.1 & 5.3 \\ \cline{2-7}
\multicolumn{2}{r|}{\textit{Median Prec}} & 5.2 & 6.4 & 3.7 & 4.0 & 6.3\\
\multicolumn{7}{r}{~}\\\hline
 & train & \cellcolor[HTML]{C0C0C0}34.0 & 9.6 & 17.9 & 20.3 & 23.7 \\  
 & farsecsq & 14.4 & 11.5 & 8.5 & 10.6 & \cellcolor[HTML]{C0C0C0}21.4 \\  
 & farsectwo & \cellcolor[HTML]{C0C0C0}26.0 & 17.9 & 13.0 & 15.5 & 15.3 \\  
 & farsec & 35.6 & \cellcolor[HTML]{C0C0C0}51.4 & 19.3 & 21.6 & 30.0 \\  
 & clni & \cellcolor[HTML]{C0C0C0}83.3 & 52.9 & 24.0 & 18.2 & 19.4 \\  
 & clnifarsecsq & 14.4 & \cellcolor[HTML]{C0C0C0}17.9 & 8.6 & 8.6 & 12.7 \\  
 & clnifarsectwo & \cellcolor[HTML]{C0C0C0}26.3 & 11.0 & 12.1 & 15.3 & 10.5 \\  
\multirow{-8}{*}{Derby} & clnifarsec & 34.0 & \cellcolor[HTML]{C0C0C0}52.8 & 16.0 & 31.9 & 23.9 \\ \cline{2-7}
\multicolumn{2}{r|}{\textit{Median Prec}} & 30.2 & 17.9 & 14.5 & 16.9 & 20.4\\
\multicolumn{7}{r}{~}\\\hline
\rowcolor[HTML]{ECF4FF}\multicolumn{2}{r|}{\textit{Overall median Prec}} & 8.9 & 6.4 & 3.7 & 4.0 & 4.9\\\hline
\end{tabular}
\label{tbl:precisionRes}
\end{table*}

\begin{table*}[!htbp]
\centering
\caption{{\bf RQ1} results: f-measure. F-measure (or f-score) is defined as the harmonic mean of the model's precision and recall. Higher values are better. Same as format as Table~\ref{tbl:resRecall}; i.e. best results are shown in grey.}
\begin{tabular}{l|l|c|c|c|c|c}
\multicolumn{2}{c}{} & \begin{tabular}[c]{@{}c@{}}Prior state\\ of the art\\ ~\cite{peters2018text}\end{tabular} & \begin{tabular}[c]{@{}c@{}}Optimize\\ learners\\ (only)\end{tabular} & \begin{tabular}[c]{@{}c@{}}Data\\ pre-processing\\ (no tuning)\end{tabular} & \begin{tabular}[c]{@{}c@{}}Data\\ pre-processing\\ (tuned)\end{tabular} & \begin{tabular}[c]{@{}c@{}}Tune both\\ (dual)\end{tabular} \\ \hline
\rowcolor[HTML]{ECF4FF} 
\multicolumn{1}{c|}{\cellcolor[HTML]{ECF4FF}\textbf{Project}} & \multicolumn{1}{c|}{\cellcolor[HTML]{ECF4FF}\textbf{Filter}} & \textbf{FARSEC} & \textbf{\begin{tabular}[c]{@{}c@{}}DE+\\ Learners\end{tabular}} & \textbf{Pre-processors} & \textbf{\begin{tabular}[c]{@{}c@{}}DE+\\ Pre-processors\end{tabular}} & \textbf{SWIFT} \\ \hline
 & train & \cellcolor[HTML]{C0C0C0}20.8 & 6.7 & 3.0 & 4.3 & 3.8 \\  
 & farsecsq & \cellcolor[HTML]{C0C0C0}18.3 & 6.2 & 2.7 & 3.5 & 5.2 \\  
 & farsectwo & \cellcolor[HTML]{C0C0C0}20.8 & 6.2 & 3.0 & 3.5 & 3.2 \\  
 & farsec & \cellcolor[HTML]{C0C0C0}20.8 & 6.6 & 2.4 & 5.0 & 5.4 \\  
 & clni & \cellcolor[HTML]{C0C0C0}20.0 & 6.8 & 3.2 & 5.5 & 3.3 \\  
 & clnifarsecsq & \cellcolor[HTML]{C0C0C0}11.9 & 5.3 & 2.9 & 3.6 & 5.3 \\  
 & clnifarsectwo & \cellcolor[HTML]{C0C0C0}20.0 & 7.0 & 2.8 & 6.0 & 4.3 \\  
\multirow{-8}{*}{Chromium} & clnifarsec & \cellcolor[HTML]{C0C0C0}20.0 & 4.6 & 2.9 & 4.5 & 3.8 \\ \cline{2-7}
\multicolumn{2}{r|}{\textit{Median f-score}} & 20.0 & 6.4 & 2.9 & 4.4 & 4.1\\
\multicolumn{7}{r}{~}\\\hline
 & train & 4.8 & 0.0 & 4.8 & \cellcolor[HTML]{C0C0C0}11.6 & 4.2 \\  
 & farsecsq & \cellcolor[HTML]{C0C0C0}4.0 & 2.6 & 2.8 & 2.9 & 2.9 \\  
 & farsectwo & \cellcolor[HTML]{C0C0C0}4.2 & 2.8 & 2.3 & 2.5 & 2.6 \\  
 & farsec & \cellcolor[HTML]{C0C0C0}8.3 & 6.5 & 3.5 & 5.8 & 6.4 \\  
 & clni & 0.0 & \cellcolor[HTML]{C0C0C0}11.1 & 5.3 & 8.6 & 7.5 \\  
 & clnifarsecsq & \cellcolor[HTML]{C0C0C0}3.0 & 2.4 & 2.4 & 2.4 & 2.7 \\  
 & clnifarsectwo & 2.8 & 2.9 & 2.6 & 2.6 & \cellcolor[HTML]{C0C0C0}3.0 \\  
\multirow{-8}{*}{Wicket} & clnifarsec & \cellcolor[HTML]{C0C0C0}9.8 & 6.5 & 4.0 & 6.5 & 6.7 \\ \cline{2-7}
\multicolumn{2}{r|}{\textit{Median f-score}} & 4.1 & 2.8 & 3.2 & 4.4 & 3.6\\
\multicolumn{7}{r}{~}\\\hline
 & train & 12.5 & \cellcolor[HTML]{C0C0C0}33.3 & 5.5 & 9.5 & 10.1 \\  
 & farsecsq & 7.4 & \cellcolor[HTML]{C0C0C0}26.1 & 3.8 & 6.4 & 7.8 \\  
 & farsectwo & \cellcolor[HTML]{C0C0C0}30.8 & 26.1 & 5.1 & 6.2 & 9.2 \\  
 & farsec & 6.3 & \cellcolor[HTML]{C0C0C0}31.6 & 5.6 & 13.3 & 8.9 \\  
 & clni & 9.5 & \cellcolor[HTML]{C0C0C0}33.3 & 8.5 & 11.3 & 9.9 \\  
 & clnifarsecsq & 16.3 & \cellcolor[HTML]{C0C0C0}28.6 & 7.9 & 13.3 & 5.2 \\  
 & clnifarsectwo & 12.9 & \cellcolor[HTML]{C0C0C0}26.1 & 8.1 & 17.1 & 8.1 \\  
\multirow{-8}{*}{Ambari} & clnifarsec & 25.0 & \cellcolor[HTML]{C0C0C0}28.6 & 12.5 & 20.7 & 9.5 \\ \cline{2-7}
\multicolumn{2}{r|}{\textit{Median f-score}} & 12.7 & 28.6 & 6.8 & 12.3 & 9.1\\
\multicolumn{7}{r}{~}\\\hline
 & train & 10.8 & \cellcolor[HTML]{C0C0C0}21.4 & 6.5 & 7.1 & 19.2 \\  
 & farsecsq & 7.9 & 9.7 & 11.4 & 14.3 & \cellcolor[HTML]{C0C0C0}14.7 \\  
 & farsectwo & 7.9 & \cellcolor[HTML]{C0C0C0}12.8 & 5.4 & 7.1 & 8.2 \\  
 & farsec & \cellcolor[HTML]{C0C0C0}11.1 & 7.5 & 4.7 & 6.4 & 10.0 \\  
 & clni & 7.5 & 10.2 & 5.4 & 7.2 & \cellcolor[HTML]{C0C0C0}12.3 \\  
 & clnifarsecsq & 6.8 & \cellcolor[HTML]{C0C0C0}14.6 & 12.2 & 10.2 & 12.6 \\  
 & clnifarsectwo & 7.0 & 7.4 & 7.0 & 7.2 & \cellcolor[HTML]{C0C0C0}9.3 \\  
\multirow{-8}{*}{Camel} & clnifarsec & \cellcolor[HTML]{C0C0C0}13.3 & 7.9 & 7.0 & 7.4 & 9.7 \\ \cline{2-7}
\multicolumn{2}{r|}{\textit{Median f-score}} & 7.9 & 10.0 & 6.8 & 7.2 & 11.2\\
\multicolumn{7}{r}{~}\\\hline
 & train & \cellcolor[HTML]{C0C0C0}36.0 & 15.8 & 26.7 & 30.0 & 35.0 \\  
 & farsecsq & 22.8 & 19.1 & 14.7 & 18.4 & \cellcolor[HTML]{C0C0C0}32.4 \\  
 & farsectwo & \cellcolor[HTML]{C0C0C0}33.6 & 27.3 & 20.2 & 25.1 & 25.6 \\  
 & farsec & 36.8 & \cellcolor[HTML]{C0C0C0}48.1 & 28.6 & 31.4 & 40.9 \\  
 & clni & 37.0 & \cellcolor[HTML]{C0C0C0}47.4 & 33.8 & 27.9 & 30.1 \\  
 & clnifarsecsq & 22.8 & \cellcolor[HTML]{C0C0C0}27.3 & 15.4 & 15.2 & 21.3 \\  
 & clnifarsectwo & \cellcolor[HTML]{C0C0C0}30.3 & 18.5 & 19.8 & 24.4 & 18.1 \\  
\multirow{-8}{*}{Derby} & clnifarsec & 36.0 & \cellcolor[HTML]{C0C0C0}48.7 & 25.3 & 40.4 & 35.2 \\ \cline{2-7}
\multicolumn{2}{r|}{\textit{Median f-score}} & 34.8 & 27.3 & 22.8 & 26.5 & 31.3\\
\multicolumn{7}{r}{~}\\\hline
\rowcolor[HTML]{ECF4FF}\multicolumn{2}{r|}{\textit{Overall median F-score}} & 12.7 & 10.0 & 6.8 & 7.2 & 9.1\\\hline
\end{tabular}
\label{tbl:f-measureRes}
\end{table*}

\subsubsection{Precision and F-Measure Results}

\responseChangeTwo{For the sake of completeness, we also provide the results of precision and f-measure. Table~\ref{tbl:precisionRes} and Table~\ref{tbl:f-measureRes} present the corresponding precision and f-measure results from each technique besides FARSEC. We make the following remarks about these results. 
\begin{itemize}
\item
The decreasing trends are expected, as we select g-measure as our optimization target, which increases the recall and  sacrifices the precision value. But, to some extent, these results also confirm the correctness of our choice. On the one hand, the improvement of recall with SWIFT is significant. On the other hand, for 4 out of 5 projects, the sacrifice of precision is moderate. For tasks such as bug report classification with the imbalanced data characteristic, as well in the context of security, in general, positive examples such as security bug reports are preferred not to be missed out. Hence, we would still recommend optimizing g-measure for future studies.
\item
There is little information gain in exploring
both precision and f-measure since these results nearly echo each other (reason: f-measure is calculated as a combination of precision and recall).
\item
We admit the importance of precision, however, in some special domains such as security, there is little information gain in exploring precision results. As seen from our results, none of the techniques (including FARSEC) performs well under the precision metric. Hence, a low precision is not necessarily a reason to ``discount'' an optimizer. When the target class is rare, such low precision might actually be expected. For example, consider a query in the Google search engine, where it takes three pages before the user finds the target page. With 10 results per page, this means that the Google search engine is scoring a precision of $\frac{1}{30} \approx 3\%$. In this case, as precision is the fraction of retrieved pages that are relevant, such low precision is only a problem of time cost since the user wastes much time exploring irrelevant results before finding the target they care about. Our IFA results in \tbl{resIFA}, from the aspect of effort, also shows that, in the case of bug report classification, these low precision results {\em do not} lead to too much wasted time (evidence: the last row of \tbl{resIFA} shows that users need to explore 15 to 22 false positives before finding a real security bug report -- which is a small number when considering the size of total bug reports).
\end{itemize}}

\subsection{RQ2}

\begin{RQ}
{\bf RQ2.} 
When learning how to  \responseChange{distinguish} security bug reports,
is it better to dual optimize the learners \responseChange{and} the data pre-processors?
\end{RQ}

\begin{table*}[!htbp]
\centering
\caption{How often is each treatment seen to be best in Table~\ref{tbl:resRecall}, Table~\ref{tbl:resFalseAlarm} and Table~\ref{tbl:resIFA}.}
\begin{tabular}{c|c|c|c}
\hline
\rowcolor[HTML]{ECF4FF} 
\textbf{Metric} & \textbf{Rank} & \textbf{Method} & \textbf{Win Times} \\ \hline
 & 1 & {\IT} & 31/40 \\  
 & 2 & Pre-processors & 14/40 \\  
 & 3 & DE+Pre-processors & 13/40 \\ 
\multirow{-4}{*}{Recall} & 4 & DE+Learners & 3/40 \\
\multicolumn{4}{r}{~}\\\hline
 & 1 & DE+Learners   & 14/40 \\  
 & 2 & Pre-processors & 1/40 \\  
 & 3 & {\IT} & 1/40 \\  
\multirow{-4}{*}{\begin{tabular}[c]{@{}c@{}}False\\ Positive \\ Rate \end{tabular}} & 4 & DE+Pre-processors & 0/40 \\
\multicolumn{4}{r}{~}\\\hline
 & 1 & DE+Learners  & 22/40 \\  
 & 2 & {\IT} & 18/40 \\  
 & 3 & DE+Pre-processors & 4/40 \\  
 \multirow{-4}{*}{IFA}& 4 & Pre-processors & 3/40 
\end{tabular}
\label{tbl:rank}
\end{table*}

This research question explores the merits of dual optimization of learner plus pre-processor versus just optimizing one or the other. To answer this question, we count how often each method achieves top-rank (and has gray-colored results) across all three metrics of the rows in Table~\ref{tbl:resRecall}, Table~\ref{tbl:resFalseAlarm} and Table~\ref{tbl:resIFA}.  

\responseChange{Those count results are shown in \tbl{rank}. From this table, we can say, in terms of recall:
\bi
\item
{\IT}'s dual optimization is clearly the best.
\item
Optimize just the data pre-processors comes a distant second.
\item
And optimize just the learners (with DE+Learners) is even worse.
\ei
Hence we say that, when distinguishing security bug reports, it is not enough to just tune the learners.}

In terms of false positive rates, we see that:
\bi
\item Optimize just the learner is a comparatively better method than other methods. 
\item Other treatments do not do well on the false alarm scale.
\ei
That said, optimize just the learner achieves a score of 14/40 -- which is not even half the results. Hence, based on false positive rates, we cannot comment on what works best for improving this metric.

In terms of IFA (initial false alarms), we see that:
\bi
\item Methods that do not optimize a learner do not perform well.
\item There is is no clear winner for the best method since DE+Learners or {\IT} perform nearly the same as each other.
\ei

Based on the above observations, we could sum up the conclusions:
\bi
\item
\responseChange{Our experiment results show} that dual optimization works well for recall. 
\item Also, not optimizing the learners performs badly for IFA.
\item There is no clear pattern in \tbl{rank} regarding false positive rates. 
\ei
That said, the results for false positive rates seen in Table~\ref{tbl:resFalseAlarm} are somewhat lower than the false positive rates seen in other software analytic papers~\cite{menzies2006data}. Hence, on a more positive note, we can still recommend dual optimization since:
\bi
\item It has many benefits (much higher recalls).
\item With no excessive cost (not large increase in false alarms; IFA results are nearly as good as other methods).
\ei

\begin{table*}[!htbp]
\centering
\caption{Average runtime (in minutes) of optimizing all learner's hyperparameters, pre-processor's hyperparameters and running {\IT}. Note that DE3 terminates after 3 generations and DE10 terminates after 10 generations.}
\begin{tabular}{c|c|c|c|c}
\hline
\rowcolor[HTML]{ECF4FF} 
\textbf{Project} & \multicolumn{1}{c|}{\cellcolor[HTML]{ECF4FF}\textbf{DE3}} & \multicolumn{1}{c|}{\cellcolor[HTML]{ECF4FF}\textbf{DE10}} & \multicolumn{1}{c|}{\cellcolor[HTML]{ECF4FF}\textbf{\begin{tabular}[c]{@{}c@{}}Data\\ Pre-processor\\ Optimization\end{tabular}}} & \multicolumn{1}{c}{\cellcolor[HTML]{ECF4FF}\textbf{SWIFT}} \\ \hline
Chromium & 455 & 876 & 20 & 12 \\ 
Wicket & 8 & 11 & 8 & 5 \\ 
Ambari & 8 & 11 & 8 & 5 \\ 
Camel & 8 & 11 & 8 & 5 \\ 
Derby & 8 & 11 & 8 & 5 \\ \hline
\end{tabular}
\label{tbl:tuningtime}
\end{table*}

\responseChange{Further to this comment of ``no excessive cost'', \tbl{tuningtime} shows the average runtime for each treatment. From the table, optimization on learners with the differential evolution algorithm consumes much more CPU time than others, while dual optimization as {\IT} shows slight advantages even better than optimizing data pre-processors. In addition, during our experiment, we also notice that, learners such as K Nearest Neighbors and Multilayer Perceptron can be slow to optimize, especially for large datasets such as the Chromium project which has about 20,000 data instances. However, since these learners are rarely selected as a ``best'' learner (see from Table~\ref{tbl:bestlearner}), we would further recommend not using those learners for bug report classification task.}

\begin{table*}[!htbp]
\centering
\caption{The time of each learner that is selected as the ``best'' learner.}
\begin{tabular}{c|c|c|c|c|c}
\hline
\rowcolor[HTML]{ECF4FF} 
\textbf{Learner} & \multicolumn{1}{c|}{\cellcolor[HTML]{ECF4FF}\textbf{FARSEC}} & \multicolumn{1}{c|}{\cellcolor[HTML]{ECF4FF}\textbf{\begin{tabular}[c]{@{}c@{}}DE+\\ Learners\end{tabular}}} & \multicolumn{1}{c|}{\cellcolor[HTML]{ECF4FF}\textbf{\begin{tabular}[c]{@{}c@{}}Data\\ Pre-processor\end{tabular}}} & \multicolumn{1}{c|}{\cellcolor[HTML]{ECF4FF}\textbf{\begin{tabular}[c]{@{}c@{}}Data\\ Pre-processor\\ Optimization\end{tabular}}} & \multicolumn{1}{c}{\cellcolor[HTML]{ECF4FF}\textbf{SWIFT}} \\ \hline
Naive Bayes & \dbox{6} & \dbox{21} & \dbox{23} & \dbox{16} & \dbox{17} \\ 
Logistic Regression & \dbox{16} & \dbox{5} & \dbox{5} & \dbox{4} & \dbox{3} \\ 
Multilayer Perceptron & \dbox{6} & \dbox{3} & \dbox{0} & \dbox{9} & \dbox{6} \\ 
Random Forest & \dbox{10} & \dbox{9} & \dbox{12} & \dbox{11} & \dbox{13} \\ 
K Nearest Neighbors & \dbox{2} & \dbox{2} & \dbox{0} & \dbox{0} & \dbox{1} \\ \hline
\multicolumn{6}{c}{\mbox{KEY: \colorbox{black!5}{\bf 5}\colorbox{black!10}{\bf 10}\colorbox{black!15}{\bf \textcolor{black}{15}}\colorbox{black!20}{\bf \textcolor{black}{20}}\colorbox{black!25}{\bf \textcolor{black}{25}} times selected.}} \\ 
\end{tabular}
\label{tbl:bestlearner}
\end{table*}

\subsection{RQ3} \label{sec:MAP}

\begin{RQ}
\responseChange{{\bf RQ3.} Can hyperparameter optimization further improve the performance of ranking security bug reports?}
\end{RQ}

\begin{figure*}[!htbp]
\centering
\includegraphics[width=10cm]{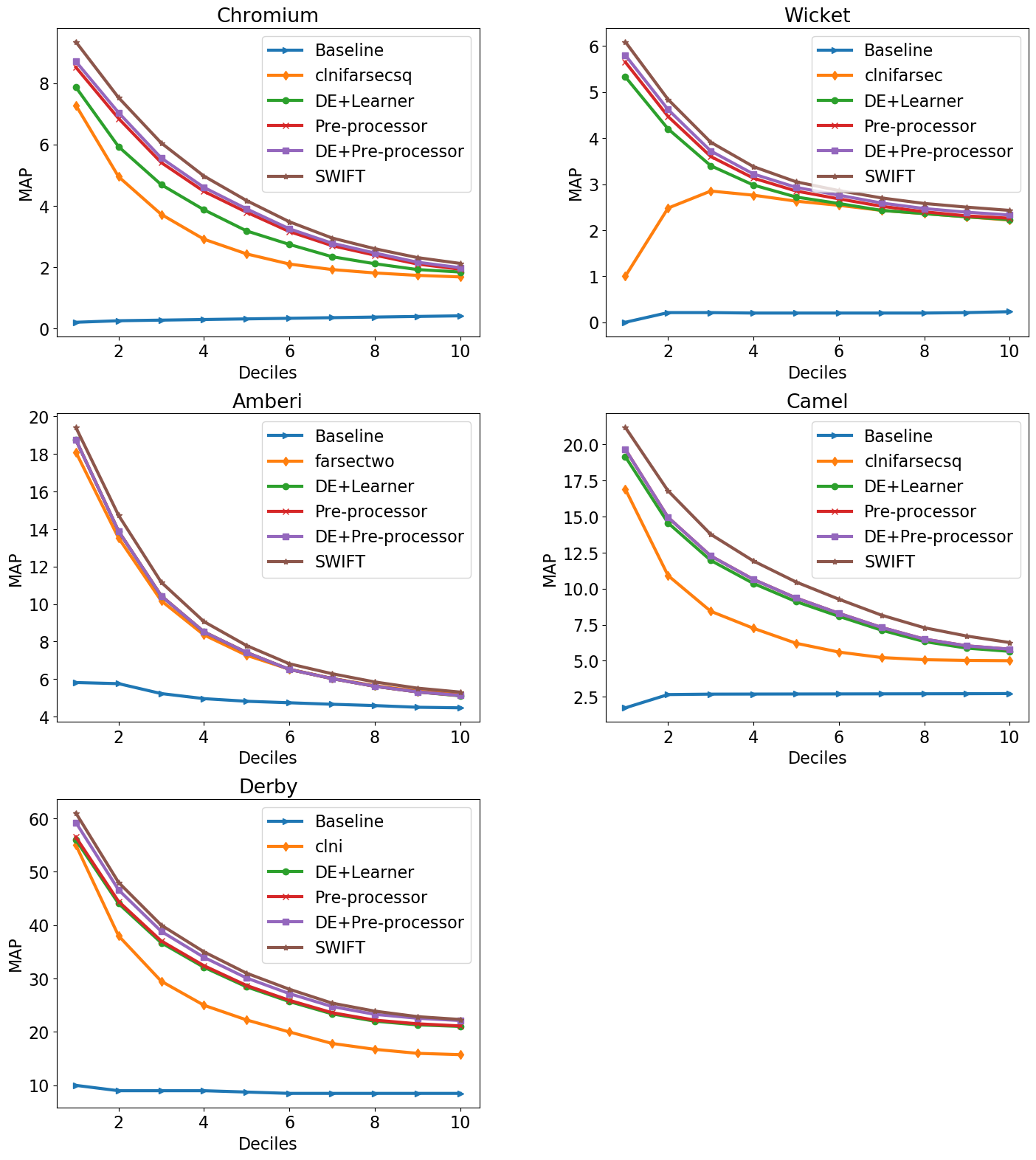}
\caption{Comparison of different treatments in ranking bug report prediction results. This plot shows its results using the deciles of \eq{four} (from \S\ref{sec:metrics}) and {\em higher} y-axis is {\em better}. Different treatments are denoted with lines of different colors. Specifically, the baseline (shown in blue color) is the method that does not apply any ranking technique (i.e., with the original chronological order). The orange line denotes the best ranking results from FARSEC among all filters.}
\label{fig:MAP}
\end{figure*}

\responseChange{As the users of the bug reports, one of the major requirements is to distinguish as many actual security bug reports as possible. Our previous treatments are trying to seek a balance between recall and specificity, as stated in Section~\ref{sec:metrics}. The result of choosing \textit{g-measure} as the optimization target is an increment of recall while at the cost of increasing false positive rate at the same time (see Table~\ref{tbl:resRecall} and Table~\ref{tbl:resFalseAlarm}). This usually could indicate that developers who use such tools would need to spend more time and effort to check those unexpected false positive predictions.}

\responseChange{For many prominent applications such as web search engine, what is germane to users is how many good results are on the first page or the first two or three pages. Inspired by this, a ranking result of predicted bug reports would therefore be more helpful and reduce the required effort for developers. As we describe in Section~\ref{tion:farsec}, FARSEC employs a ranking method that sorts the predicted security bug reports. As a result, the actual security bug reports are closer to the top of the rank list.}

\responseChange{We apply the same ranking technique as FARSEC, while the learners and/or pre-processors are optimized. The evaluation results based on the MAP metric are shown in Figure~\ref{fig:MAP}. In the figure, the baseline (shown in blue color) is the method that does not apply any ranking technique (i.e., with the original chronological order). The orange line denotes the best ranking results from FARSEC among all filters. The other treatments are denoted with lines of different colors.}

\responseChange{The key observations from the figure are:
\bi
\item The baseline method performs badly (the blue line) since this is with no ranking technique, whatsoever.
\item In all data sets, the ranking generated using the prior
state-of-the-art (the orange line for FARSEC filters) is below other treatments that try to rank the predicted bug reports.
\item In a result that is consistent with the main message of this paper, in all data sets, the rankings generated by dual optimization (the brown {\IT} line) is above other methods.
\ei
}

\responseChange{The experiment results of ranking security bug reports, as well as results in previous research questions, could indicate that our proposed dual optimization of learners and pre-processors are promising. This approach could be recommended to better aid practitioners with similar domain tasks.}

\section{Discussion}
\label{discussion}

\responseChangeTwo{SWIFT has demonstrated new results that improve the prior state-of-the-art. Speaking more broadly, what are the other lessons that could be taken from this work? We make the following comments.}

\responseChangeTwo{Firstly, at the general application level, we have shown here it is possible to reason about rare event data (e.g., here the target security bug reports can be as rare as only taking up 1\% of the total bug reports). Apart from the security case studied here, another lesson we would offer is that (sometimes) practitioners do not need (much) data to start data mining. This is an intriguing statement, since in this era of ``big data'', it is often assumed that scare of data would be a large obstacle. Here we offer a somewhat more optimistic comment: \textit{effective models can be built even when data is scarce}.}

\responseChangeTwo{Secondly, at the methodological level, we offer the following suggestion: {\em avoid using AI tools ``off-the-shelf'' without modifying them for the local domain}.
SE practitioners need to develop specialized machine learning tools that are better suited to particular SE problems. 
Existing machine learning algorithms that we might call ``general AI machine learning tools'' maybe not ``general'' at all. Rather, they are tools whose default settings were chosen according to the data used {\em in the past} to commission those tools. Hyperparameter optimization tools should always be applied to adjust AI tools to the local data.}

\responseChangeTwo{(Aside: One objection to the above point is that such optimization process can be unduly expensive. This objection is certainly true when we use traditional hyperparameter optimizers (e.g. genetic algorithms that evaluate thousands to millions of options~\cite{holland1992genetic}).
However, our empirical results from Table~\ref{tbl:tuningtime} shows that effective hyperparameter optimization can be accomplished in minutes. We note that, aside from data mining for security, previous researchers have achieved similar ``fast optimization'' results in several other SE domains~\cite{agrawal2019dodge}.)}

\responseChangeTwo{We are not the only researchers who make this second point. Other researchers in the software analytics literature also advocate tuning general AI tools to SE tasks. For example, Binkley et al.~\cite{binkley2018need} note that information retrieval tools for SE often equate word frequency with word importance, even though the number of occurrences of a variable name such as ``tmp'' is not necessarily indicative of its importance. They argue that the negative impacts of such differences manifest themselves when ``off-the-shelf'' information retrieval tools are applied in the software domain. Another example comes from sentiment analysis. Standard sentiment analysis tools are usually trained on non-SE data (e.g., the Wall Street Journal or Wikipedia). Novielli et al.~\cite{novielli2018benchmark} recently developed their own sentiment analysis for the software engineering domain. After re-training those tools on an SE corpus, they found not only better performance at predicting sentiment, but also more agreement between different sentiment analysis tools.}

\responseChangeTwo{Thirdly, it is natural to ask whether optimizing
data pre-processors is more important than optimizing the learners (or vice versa). In reply, we say that there is no evident hints from our empirical results show that one of them has obvious advantages over the other. In fact, recalling {\bf RQ2}, we say that (at least in this domain) it is better to tune \textit{both}.}

\responseChangeTwo{Fourthly, another question we are asked is ``in other domains, do our results say that some learners/pre-processors will perform better?''. Our results do not support such conclusion. Table~\ref{tbl:bestlearner} shows that the ``best'' classifier is highly variable across our datasets. Hence, we cannot offer one general conclusion for all projects. However, what we do offer
is a general method for finding the best local solution. Further, as shown by the runtime in Table~\ref{tbl:tuningtime}, it may not be especially slow to apply our  general method for finding the best local solution. }

\section{Threats to Validity}
\label{threats}

As to any empirical study, biases can affect the final results. Therefore, conclusions drawn from this work must be considered with threats to validity in mind. 

\textbf{Sampling Bias.} Sampling bias threatens any classification experiment. For example, the data sets used here come from FARSEC, i.e., one Chromium project and four Apache projects in different application domains. In addition, the bug reports from Apache projects are randomly selected with a BUG or IMPROVEMENT label for each project with extra labeling effort.
 
\textbf{Learner Bias.} Research into automatic classifiers
is a large and active field. While different machine learning algorithms have been developed to solve different classification problem tasks.
Any data mining study, such as this paper, can only use a small subset of the known classification algorithms. For this work, we selected our learners such that we can compare our results to prior work. Accordingly, we used the same learners as Peters et al. in their FARSEC research.

\textbf{Input Bias.} Our results come from the space of hyperparameter optimization explored in this paper. In theory, other ranges might lead to other results. That said, our goal here is not to offer the {\em best} optimization but to argue that {\em dual} optimization of data pre-processors and learners is preferable to optimize either, just by itself. For those purposes, we would argue that our current results suffice.

\responseChange{\textbf{Evaluation Bias.} In our work, we choose some commonly used metrics as FARSEC for evaluation purpose and set \textit{g-measure} as our optimization target. We do not use some other metrics because relevant information is not available to us or we think they are not suitable enough to this specific task (e.g., precision). In addition, we use equal weight in recall and specificity in the definition of g-measure, which is widely adopted in existing literature. We agree that it is important for these two elements to be re-weighted for different tasks, and this can be further explored as one of our future directions. Our implementation is flexible and we can adjust to proper metrics or balances with minor code modification.}

\section{Conclusion}
\label{conclusion}

\responseChange{Distinguishing security bug reports from other kinds of bug reports is a pressing problem that threatens not only the viability of software services, but also consumer confidence in those services. Prior results on how to distinguish security bug reports have had issues with the scarcity of target data (specifically, such incidents occur rarely). In a recent TSE'18 paper, Peters et al. proposed some novel filtering algorithms to help improve security bug report classification. Results from FARSEC show that such filtering techniques can improve the performance.}

But more than that, our experiments show that we can further do better than FARSEC using hyperparameter optimization of data mining learners and data pre-processors. Our results show that it is more advantageous to apply {\em dual} optimization of {\em both} the data-processor {\em and} the learner, which we will recommend in solving similar problems in future work.

\begin{acknowledgements}
This work was partially funded
via an NSF-CISE grant \#1909516.

\end{acknowledgements}

\bibliographystyle{spbasic} 
\bibliography{main.bbl} 


\end{document}